\documentclass[a4paper,11pt]{article}
\pdfoutput=0 % if your are submitting a pdflatex (i.e. if you have
\usepackage{jinstpub} % for details on the use of the package, please
\usepackage{savesym}
\usepackage{amsmath}
\savesymbol{iint}
\savesymbol{iiint}
\usepackage{txfonts}
\restoresymbol{TXF}{iint}
\restoresymbol{TXF}{iiint}

\usepackage{graphicx, subfigure}
\usepackage{lscape}
\usepackage{epsfig}
\usepackage{longtable}
\usepackage{amssymb}
\usepackage{placeins}
\usepackage{multirow}
\usepackage{epstopdf} 
\usepackage{multirow}
\usepackage[flushleft]{threeparttable}
\usepackage{array}

\usepackage{rotating}
\usepackage{chngpage}
\usepackage{booktabs}

\usepackage{lineno}
\usepackage{lineno}
\usepackage{multirow}
\usepackage{booktabs}
\usepackage{amsbsy}
\usepackage{amsmath}
\usepackage{color}

\title{\boldmath Very High-Energy Gamma-Ray Follow-Up Program Using Neutrino Triggers from IceCube}

\emailAdd{Dariusz.Gora@desy.de, Elisa.Bernardini@desy.de,
Robert.Franke@desy.de}

\collaboration{The IceCube Collaboration:}
\author[2]{M.~G.~Aartsen,}
\author[34]{K.~Abraham,}
\author[52]{M.~Ackermann,}
\author[16]{J.~Adams,}
\author[12]{J.~A.~Aguilar,}
\author[30]{M.~Ahlers,}
\author[42]{M.~Ahrens,}
\author[24]{D.~Altmann,}
\author[32]{K.~Andeen,}
\author[48]{T.~Anderson,}
% ...

\abstract{We describe and report the status of a neutrino-triggered program in IceCube that generates real-time alerts for gamma-ray follow-up observations by 
atmospheric-Cherenkov telescopes (MAGIC and VERITAS). While IceCube is capable of monitoring the whole sky continuously, high-energy gamma-ray telescopes have 
restricted fields of view and in general are  unlikely to be observing a potential neutrino-flaring source at the time such neutrinos are recorded. 
The use of neutrino-triggered alerts thus aims at 
increasing the availability of simultaneous multi-messenger data during potential neutrino flaring activity, which can increase the discovery 
potential and constrain the phenomenological interpretation of the high-energy emission of selected source classes (e.g. blazars). 
The requirements of a fast and stable online analysis of potential neutrino signals and its operation are presented, along  with first results of the  program operating between   14 March 2012  and   31  December 2015. }


\author [51a] {M.~L.~Ahnen,}
\author [52a] {S.~Ansoldi,}
\author [53] {L.~A.~Antonelli,}
\author [54] {P.~Antoranz,}
\author [55] {A.~Babic,}
\author [56] {B.~Banerjee,}
\author [57] {P.~Bangale,}
\author [57,72] {U.~Barres de Almeida,}
\author [58] {J.~A.~Barrio,}
\author [59,73] {J.~Becerra Gonz\'alez,}

% ...

\begin{document}
\maketitle
\flushbottom

\section{Introduction} 
\label{intro}
Observations of astrophysical neutrinos may help to answer some of the
most fundamental questions in astrophysics, in particular the mystery
of the source of cosmic rays (for a general discussion see
\cite{uhecr}). For neutrinos in the TeV range, prime source candidates
are Galactic supernova remnants \cite{remnats}. Neutrinos in the PeV
range and above are suspected to be produced by active galactic nuclei
(AGN) and gamma ray bursts (GRB) with many AGN models predicting a
significant neutrino flux
\cite{Atoyan:2004pb,PhysRevD.80.083008,Mucke2003593}. However, the recent results from  the IceCube Collaboration  strongly disfavor GRBs as sources of the highest energy cosmic rays ~\cite{grb_icecube}. Recently, the IceCube Collaboration has  also reported the first observation of a cosmic diffuse
neutrino flux which lies in the 100 TeV to PeV range~\cite{science_icecube}. Individual neutrino sources, however,
could not be identified up to now. While many astrophysical sources of
origin have been suggested \cite{icecube_inter}, there is not yet
enough information to narrow down the possibilities to any particular
source or source class. 

Gamma-ray observations  by imaging atmospheric-Cherenkov telescopes (IACTs) such as VERITAS~\cite{veritas}, HESS~\cite{hess}  or MAGIC~\cite{magic} have also a potential to find hadronic $\gamma$-rays from the neutrino directions and to identify neutrino sources~\cite{santander,schussler}. The expected neutrino  flux  from observed high-energy gamma-ray fluxes of blazars in their  brightest states (e.g.~the flares of Markarian 501 in
1997~\cite{aha1} and 2005~\cite{aha10}, Markarian 421 in 2000/2001~\cite{aha2} and
2008~\cite{mrk421} and PKS 2155-304 in 2006~\cite{aha3}) can be at the level 
of the neutrino flux  detected by the high-energy neutrino telescopes
~\cite{n_mrk421,pssource,pssourcetime}.

%Recent results obtained by the IceCube
%Collaboration~\cite{pssource,pssourcetime} indicate that high-energy
%neutrino telescopes have reached a sensitivity to neutrino fluxes
%comparable to the observed high-energy gamma-ray fluxes of blazars in their  %brightest states (e.g.~the flares of Markarian 501 in
%1997~\cite{aha1} and 2005~\cite{aha10}, Markarian 421 in 2000/2001~\cite{aha2} and
%2009~\cite{mrk421} and PKS 2155-304 in 2006~\cite{aha3})
%i.e. the produced neutrino flux from these objects 
 %is close to the sensitivity of current generation neutrino telescopes, 
% under the assumption that any associated neutrino emission exhibits a flux
%enhancement comparable to that which is observed in gamma rays.
%, it
%should now be possible to measure statistically significant evidence
%for neutrino flares during such high states.
%
%The detection of cosmic neutrinos by high-energy neutrino telescopes
%is very challenging, due to the small neutrino interaction
%cross-section and the large background of atmospheric neutrinos.
%Simultaneous measurements using neutrino and electromagnetic
%observations (the so-called ``multi-messenger'' approach) can increase
%the chances of identifying a source of astrophysical neutrinos, by
%reducing the trial factor statistical penalty arising from observation of multiple
%sky regions over different time periods.

For sources which manifest large time variations in the emitted
electromagnetic radiation, the signal-to-noise ratio can be increased
by searching for periods of enhanced neutrino emission (a
time-dependent search).  Of special interest is the relation of these
periods of enhanced neutrino emission with periods of strong
high-energy gamma-ray emission. However, IACTs have a duty cycle limited to the clear, dark nights (roughly 10\% of total time), such correlation studies are not always possible after the neutrino flare has occurred. Therefore, it is
desirable to ensure the availability of simultaneous neutrino and
high-energy gamma-ray data for periods of interest. This can be
achieved by an online neutrino flare search that alerts the partner
IACT experiments when an elevated rate of neutrino events from the
direction of a source candidate is detected. Such a Neutrino Triggered
Target of Opportunity program (NToO), using a list of pre-defined
sources, has been developed and has been operating since 2006 using
the AMANDA neutrino telescope to initiate gamma-ray follow-up
observations by MAGIC~\cite{icrc2007ps:NToO,icrc2o15:NToO}. 

Similarly, one can conduct a search for neutrinos from short transient
sources (with a time scale of 100 seconds), such as gamma-ray bursts
(GRBs) (see e.g. \cite{wb}) and core-collapse supernovae (SNe) (see
e.g. \cite{ab}). These sources are most accessible in X-ray and
optical wavelengths, where one can observe the GRB afterglow or the
rising SN light curve, respectively. Similarly to IACTs, the field of view
and observation time of X-ray and optical telescopes are limited. Since
identification of a GRB or SN is only possible within a certain time
range (a few hours after a GRB and a few weeks after a SN explosion),
it is important to obtain electromagnetic data within these time
frames. To accomplish this, a NToO program triggering optical and X-ray
follow-up of short neutrino transients has been operating since 2008
\cite{ofu,ofu2}. Upon observing multiplets of neutrino candidates (at
least two within 100 seconds and within 3.5$^{\circ}$ (angular resolution),
from any direction) alerts are sent to the Robotic Optical Transient
Search Experiment (ROTSE) \cite{rotse}\footnote{ROTSE was used from
  2008-2014 and it is not operational anymore.} and the Palomar
Transient Factory (PTF) \cite{ptf} for optical observations, and to
Swift \cite{swift,swift2} for X-ray follow-up, depending on the
multiplet's significance. 

{
We present here a refined and enhanced implementation of the NToO system
 using the  IceCube Neutrino Observatory  (see also~\cite{rfrankethesis}).
An important goal of this program was to establish and to test procedures to trigger promptly the gamma-ray community to collect high-sensitive VHE data from specific sources during periods of time when IceCube measures a potential increase in their neutrino flux. The program is based on a multi-step neutrino selection that is applied
online at the South Pole. An alert is sent to the partner telescopes MAGIC and VERITAS in the  case that
a statistically significant cluster of neutrinos is observed from any of the monitored sources. If
the source were to be found in an enhanced flux state by the IACT follow-up observation, the
combination of the neutrino observation and the very high-energy gamma-observation could help to  establish the discovery of neutrino point sources. Furthermore, combining the two observations would
increase the potential insight into the physical processes in the source that lead to the flare.

The structure of the paper is the following: Section 2 defines  the sources used in the NToO system. Section 3 focuses on the short description of  the IceCube and IACTs detectors and their detection
principle. In Section 4 the NToO neutrino event selections and the properties of the final neutrino sample are shown.  
Section 5  describes how the significances of neutrino clusters in the NToO are calculated. Sections 6, 7, 8 describe the technical design and implementation of the NToO system. In Section  9 we present first  results of the  program operating between   14 March 2012  and   31  December 2015. In Section 10 recent and upcoming improvements of the NToO system are discussed. Finally, in Section 11
 a short summary is given.}
% The scope of the paper is following. First, we review ... Second ...
% Then, we present ... Finally, we present the results .

\section{Selection of target sources}
\label{sourcelist}

The probability of discovering extraterrestrial neutrino point sources
varies with the phenomenology of the accelerators and of their
emission mechanisms. Coincident observation of gamma rays and neutrinos
might be possible for sources where charged and neutral mesons are
produced simultaneously from hadronic p-p or p-$\gamma$
interactions. These hadronic processes may be present in variable
extragalactic objects such as BL Lacs  or flat-spectrum radio quasars (FSRQs), as well as in Galactic systems like microquasars and
magnetars. 

{
Blazars, a subset of radio-loud active galactic nuclei with relativistic jets pointing towards the Earth~\cite{padowani} are commonly classified based on the properties of the
spectral energy distribution (SED) of their electromagnetic emission. The blazar SED features two distinctive peaks: a low-energy peak between IR and X-ray energies, attributed to synchrotron emission of energetic electrons, and a high-energy peak at gamma-ray energies, which can be explained by several and possibly competing interaction and radiation processes of high-energy electrons and high-energy nuclei \cite{bottcher}. It has been suggested that blazar SED follow a sequence \cite{fossati,Dermer,cavaliere}
in which the peak energy of the  synchrotron emission spectrum decreases with increasing  blazar luminosity. Accordingly, blazars can be classified into low synchrotron peak (LSP), intermediate synchrotron peak (ISP) and high synchrotron peak (HSP) objects, a classication scheme introduced in \cite{abdo}. A second classifier is based on the prominence of emission lines in the SED over the non-thermal continuum emission of the jet. FSRQs show Doppler-broadened optical emission lines \cite{stickel}, while in the BL Lac objects the emission lines do not exist, or are hidden in a strong continuum emission.}

The probability for detection of an individual AGN neutrino
flare can be estimated based on the
predictions of different mechanisms for the observed electromagnetic
emission at high energies~\cite{Atoyan:2004pb,PhysRevD.80.083008}.  A
common feature of several models is that the class of high-energy
peaked HSP is expected to have lower gamma 
luminosity as compared to low-energy peaked LSP and FSRQs, if the observed high-energy gamma-ray emission is largely the result of interactions
of protons with ambient or self-produced radiation. With high target
matter density, the neutrino yield can be highest when the very
high-energy gamma-ray emission is strongly attenuated by internal
absorption (although the cutoff energy is somewhat uncertain). In the
case of pp-dominated scenarios, the conclusions are
different~\cite{PhysRevD.80.083008}, favoring LSP to FSRQs. In all
cases, the availability of simultaneous information on high-energy
gamma-ray emission and neutrinos is crucial  for distinguishing between
different production models.

The most interesting targets for gamma-ray follow-up observations
triggered by IceCube alerts are promising sources of TeV neutrinos,
which are either known to exhibit a bright GeV flux in gamma rays and
show extrapolated fluxes detectable by IACTs, or are already detected
by IACTs and are variable.  We consider two different target source
lists. One list was selected based on the the second {\it Fermi}-LAT
point-source catalog~\cite{fermilist}~\footnote{The third {\it Fermi}-LAT
  point-source catalog~\cite{fermilist2} or catalog of hard {\it Fermi}-LAT
  sources~\cite{2fhl} would have been more suitable, but it was not available when the selection criteria were  established and the program started.}. The following criteria were applied:
\begin{itemize}
\item Redshift $<0.6$~\footnote{The MAGIC and VERITAS telescopes have recently  detected sources with $z \sim 0.94$, PKS 1441$+$25 \cite{pks1441,pks1441veritas} and  B0218+357 \cite{b0218}); therefore  this  selection criterion (''cut'')   
will be extended to $z=1$ for the next IceCube observing season 2016/2017.}
\item {\it Fermi}-LAT variability index $>41.64$ (corresponding to the $99\%$ confidence
    level of the source being variable, see~\cite{fermilist} for  the definition of this quantity)
\item Power-law spectral index as observed with {\it Fermi}-LAT $<2.3$ (BL Lacs only~\footnote{As  shown in~\cite{PhysRevD.80.083008} for pp-scenario only BL Lacs with the spectral index below 2.2 - 2.3 are promising candidates for IceCube detection, see Figure 2 in this paper.})

\item {\it Fermi}-LAT flux [$1-100$GeV ]$>1 \cdot 10^{-9} \mbox{ph}\; \mbox{cm}^{-2}\;
\mbox{s}^{-1}$ (BL Lacs only)
\item {\it Fermi}-LAT flux $[0.1-1$GeV ]$>7 \cdot 10^{-8} \mbox{ph}\; \mbox{cm}^{-2}\;
\mbox{s}^{-1}$ (FSRQs only~\footnote{As  shown in~\cite{Atoyan:2004pb} for $p$-$\gamma$-scenario only FSRQs can be effective for interpretation of gamma-ray fluxes up to GeV energies.})
\end{itemize}
These selection criteria result in $21$ sources on the list in total
(three FSRQs and 18 BL Lacs).  This list of target sources was
combined with lists provided by the partner experiments (currently
MAGIC and VERITAS) covering the Northern Hemisphere (declination $\delta >
0^{\circ}$). All known potentially variable VHE  sources and all
sources in the {\it Fermi}-LAT monitored-sources list~\cite{fermi2} with
declination larger than $0^{\circ}$ are used. In total $109$ sources
are included in the follow-up program for the 2012/2013 IceCube
season, see Table~\ref{tab:sources2012-2013} in Appendix. As we can see from this table
43/(31) sources are present only in the VERITAS/(MAGIC) list and 35 sources are in the list for both experiments. From   November 2013 to December 2015,  the number of  sources  were reduced ~\footnote{For MAGIC,  the number of sources was reduced 
  in order  to fit to amount of observation time that was granted by MAGIC time allocation committee. } to 83 i.e. 5 for MAGIC;  65 for  VERITAS,  and 13 sources are present in both lists, see Table~\ref{tab:sources2014-2015} in Appendix. In principle,  the neutrinos could also come from unknown sources
anywhere in the sky. However, such an all-sky search was not feasible at the time the program was started due to limiting computing resources at the South Pole. 
Furthermore, an all-sky search suffers from large trial factors compared to the pre-defined source list search.

\section{The IceCube detector and IACT partners}
\label{icecube_detector}
The IceCube Neutrino Observatory \cite{icecubedetector,dompaper,pmtpaper}   is located at the geographic South Pole and was
completed at the end of 2010. The goal of the detector is to serve as
a neutrino telescope, allowing observations of neutrinos of
astrophysical origin in the TeV and PeV energy range.  Cherenkov light produced by
the secondary leptons from neutrino interactions in the vicinity of
the detector is used to detect these neutrinos.  IceCube is also
sensitive to downward-going high-energy muons and to neutrinos
produced in cosmic-ray-induced air showers.  These events represent a
background for most IceCube analyses.

For the studies presented here, only events produced by charged-current interactions of muon neutrinos are considered, because of the long range of the secondary muons which
allows for reconstructing the arrival direction of these events with
good accuracy. The pointing information relies on the secondary muon
direction, which at energies above a TeV differs from the original
neutrino direction by less than the angular resolution of the
detector.

%As the computing resources at the South Pole are limited, different
%types of software triggers are applied to lower the data event rate.
%The most important for the purposes of this work is the {\it Simple
%Multiplicity Trigger} (SMT8) which requires eight triggered optical
%modules (i.e.~four  coincidence pairs) anywhere in the detector
%within 5 $\mu s$. Most of the events which are selected by this
%trigger are composed of muons produced by cosmic rays in the
%atmosphere above the detector (about 2.5 kHz at trigger level in the
%86-string configuration).  As the data volume produced at trigger
%level is still too large to be transferred via satellite, a first
%selection has to be applied directly at the South Pole by using the
%so-called {\it Muon Filter}.  This filter aims to select
%well-reconstructed muon tracks  of any direction, i.e. from the full sky.

% and reduce the data rate to about 45 Hz. 
 {
The program presented in this work sends alerts  to IACTs for follow-up observations.
% Here we  briefly describe  how these instruments work and give short  information about the %MAGIC and VERITAS telescopes. 
IACTs detect photon-induced air showers by means of the Cherenkov light from the highly relativistic charged particles in the  shower. Due to the interplay between the emission geometry and the altitude dependent index of refraction, the Cherenkov light flash ($ \sim 10$ ns duration)  is mainly concentrated in a light pool with a radius of $\sim120$ m (for gamma or electron showers) on the ground. 
A telescope located inside the light pool can reflect the light into a 
PMT  camera. Cherenkov images of the showers are used to differentiate between gamma-ray signal and background, and to reconstruct the energy and the incoming direction of the gamma rays. 

The MAGIC telescope array is located  on the Roque de
los Muchachos Observatory (28.8$^\circ$~N, 17.9$^\circ$~W; 2200~m
above sea level), at the Canary Island of La Palma (Spain).
The MAGIC array consists of two telescopes, placed 85 m
apart, each with a primary mirror of 17 m diameter.
 The MAGIC telescopes,  with a field of view of 3.5$^{\circ}$, are able to detect cosmic gamma rays in the range 50 GeV-50 TeV. The latest performance of MAGIC is reported in \cite{magicperf}.

VERITAS is an array of four 12-m diameter imaging
atmospheric-Cherenkov telescopes located at the Fred Lawrence Whipple
Observatory  in southern Arizona ($31^{\circ}40'$N,
$110^{\circ}57'$W) at an altitude of 1.3~km. Each of the individual telescopes have  a $3.5^{\circ}$ field of view. 
Full details of the VERITAS instrument performance and sensitivity are given in \cite{VERITASperf}. }

% At this level the  basic track reconstructions are performed, mainly likelihood based with the exception of {\it Linefit}, which is an algorithm used as a seed for more precise and CPU intense reconstructions to follow. These likelihood based fits use the photon arrival time distribution for the track reconstruction~\cite{AMANDATrackReco}.
% The multi-photoelectron (MPE) likelihood function, which uses time and amplitude information of the PMT pulses, is applied after several iterations of the single photoelectron (SPE) likelihood fit that uses only the pulse leading edge time. The MPE algorithm includes a probability distribution function (PDF) that describes the scattering of photons in the ice, and is fully described in~\cite{AMANDATrackReco}.
% Track fits based on the MPE likelihood achieve a significantly improved angular resolution compared to fits based on the SPE likelihood.

%While the {\it Muon Filter} provides a sample of neutrino candidate events it is still heavily background %dominated (compared to the atmospheric neutrino rate of about 10 mHz  at the trigger level). In order to %apply further cuts with a high signal efficiency, more elaborate reconstructions with an improved angular %resolution are needed. 
%The combination of additional reconstructions and event selection cuts is referred to as the {\it Online %Level~2 Filter}.

\section{Neutrino event selection}
\label{neutrino_selection}

This section describes the online neutrino selection that is the basis for the NToO system.
As the computing resources at the South Pole are limited, different types of software triggers are applied to lower the data event rate. The most important for the purposes of this work is the {\it Simple Multiplicity Trigger} (SMT8) which requires eight triggered optical
modules (i.e.~four  coincidence pairs) anywhere in the detector
within 5 $\mu s$. Most of the events which are selected by this
trigger are composed of muons produced by cosmic rays in the
atmosphere above the detector (about 2.5 kHz at trigger level in the
86-string configuration).  As the data volume produced at trigger
level is still too large to be transferred via satellite, a first
selection has to be applied directly at the South Pole by using the
so-called {\it Muon Filter}.  This filter aims to select
well-reconstructed muon tracks  of any direction, i.e. from the full sky.

 The event selection takes place in several steps, called  ``levels`` in IceCube.
The {\it Muon Filter}  constitutes the first filtering level. It is a standard IceCube filter
and not specific to the program presented here. The subsequent {\it  Online Level 2 Filter}
is based on the input from the {\it Muon Filter} and was specifically developed to enable online analyses. Currently the {\it  Online Level 2 Filter} forms the basis of the optical and X-ray follow-up program and the NToO
system presented in this work. Based on cut variables calculated
from the {\it  Online Level 2 Filter}, an online neutrino event selection was implemented.
The main goal is to achieve a high efficiency for valid neutrino events with the
 highest  possible rejection of background.

\subsection{Muon Filter}

The {\it Muon Filter} event-selection algorithm is the basis for many
standard IceCube muon-neutrino analyses, e.g.~the searches for
neutrino point sources, searches for neutrinos from gamma-ray bursts
and measurements of the atmospheric muon-neutrino flux. The input to
the {\it Muon Filter} is all of the events that trigger the SMT8. All
events triggering the SMT8 are reconstructed with the \textit{Linefit}
first-guess algorithm as described in~\cite{AMANDATrackReco}.  The
result from this track fit forms the input to a single-photoelectron
(SPE) likelihood fit~\cite{EnergyReco,AMANDATrackReco}, which uses only the time and charge of the first
hit on each DOM. The {\it Muon Filter} decision is based on variables
calculated from the SPE fit.

\begin{table}[t]
  \centering
    \begin{tabular}{|c|c|c|}\hline
  IceCube data taking period    &  Start  & End \\
      \hline
      IC-2011& 2011 May 13             & 2012 May 16 \\
      IC-2012& 2012 May 16             & 2013 May 3 \\
      IC-2013& 2013 May 3               & 2014 May 5 \\
      IC-2014& 2014 May 5               & 2015 May 18 \\
      IC-2015& 2015 May 18             & 2016 May 20 \\ 
      
      \hline

    \end{tabular}
    \caption[Neutrino selection cut efficiency]{Summary of  IceCube data taking periods (seasons) used by NToO searches }
    \label{tab:ntoo:seasons}
\end{table}

The track reconstructions and cuts applied in the {\it Muon Filter}
have been stable over several years. However, improvements to
reconstruction algorithms, changes in the available satellite transfer
bandwidth, or changes in the data serialization format lead to small
adjustments from season to season. The IceCube seasons important for this work are  listed in the  Table~\ref{tab:ntoo:seasons}.
 For example, in the IC-2012
season an improved {\it Linefit} algorithm was used, which uses
 a linear fit with reduced weights for outliers~\cite{LinefitIMP} .
	
The {\it Muon Filter} divides the sky into two regions in which
different selection techniques are applied to remove background
events. In the first region (defined by the zenith angle $\theta \geq
{78.5}^{\circ}$) the background events are down-going muons
mis-reconstructed as up-going (or slightly above the horizon), which in
fact originate from cosmic-ray-induced air showers. The main discriminants to
remove these events are parameters characterizing the reconstruction
quality of the event. In the second region (zenith angle
$\theta<78.5^{\circ}$) both signal and background events have the same
signature, namely high-energy muon tracks. As the energy spectrum of
muons in cosmic-ray air showers ($\Phi(E) \sim E^{-3.7}$) is much
steeper than the expected signal spectra, cuts on energy-related
variables are an efficient way to reduce this background.  However, as
the current NToO is only implemented for zenith angles
$\theta>90^{\circ}$ the event-selection cuts in the second region
will not be described in detail.

In the  first region  the
{\it Muon Filter} uses a cut variable derived from the value of the
likelihood of the SPE track fit.  The definition of the cut variable
is similar to the scaled log-likelihood of the fit. All events which
are reconstructed with a zenith angle $\theta_{\text{SPE}} \geq
78.5^{\circ}$ and that fulfill
\begin{equation}
-\log_{10}  (\max\mathcal{L}_{\text{SPE}})/(N_{\text{DOM}} - 3) \leq 8.7,
\end{equation}
where $\max \mathcal{L}_\text{SPE}$ is the maximum value of the likelihood function of the SPE track fit  and $N_{\text{DOM}}$ denotes the number of triggered
DOMs in that event, passed the filter. The efficiency of the {\it Muon
  Filter} for atmospheric neutrinos is about 87\% with respect to SMT8. Neutrinos following
a spectrum of the form $\Phi(E) \sim E^{-2}$ are selected with an
efficiency of approximately 93\% with respect to SMT8. 
The cuts remained unchanged between the different IceCube
seasons i.e. from  IC-2011 to IC-2014 .  The total
event passing rate of the {\it Muon Filter} amounts to approximately
45 Hz, out of which about 18 Hz consists of events reconstructed with
zenith angle $\theta_{\text{SPE}} > 90^{\circ}$.

%{\it  {\bf Comment from MarkusV:} Later, you use the term Muon Filter. It would be best to define it
%already in this section. Maybe there is a paper that desribes what Muon
%Filter is and what it does that we can refer to? Maybe in some PS paper?
%
%I would add the term 'zenith angle', so: $\theta > 80$ deg -> zenith angle
%$\theta > 80 $deg
%}

%As we already mentioned the basis for the neutrino event selection is an on-line filter that searches for
%high-quality muon tracks.
%The full-sky rate of this filter is about $40\,$Hz for IceCube in its
%2012/2013 configuration with $86$ deployed
%strings. This rate is strongly dominated by atmospheric muons.
%In order to efficiently select neutrino events from this sample
%several elaborate reconstruction algorithms have to be applied.
%However, as the computing resources at the South Pole are limited,
%this is only possible at a lower event rate. 

\subsection{Online Level 2 Filter}

While the {\it Muon Filter} provides a sample of candidate neutrino
events it is still heavily background-dominated (compared to the
atmospheric-neutrino rate of about {10}{ mHz} at the trigger
level). In order to apply further cuts with a high signal efficiency,
more elaborate reconstructions with an improved angular resolution are
needed. As an example, the multi-photoelectron (MPE) likelihood
function, which uses the temporal and amplitude information of the PMT
pulses, is applied after several iterations of SPE likelihood fit. The
MPE algorithm includes a probability distribution function (PDF) that
describes the scattering of photons in the ice, and is fully described
in~\cite{AMANDATrackReco}.  Further reconstructions that  estimate
the angular reconstruction uncertainty are also helpful for subsequent
analyses. This combination of additional reconstruction and event-selection cuts is referred to as the {\it Online Level 2 Filter}.

The SPE fit used as an input to the {\it Muon Filter} has limited
angular resolution compared to an MPE fit.  During the first season of
running the IceCube  in its full 86-string configuration (IC-2011), the limited CPU resources at the South Pole
prohibited applying more resource-intensive reconstructions to all
events that passed the {\it Muon Filter}. Therefore, it was necessary
to apply event-selection cuts to the events passing the {\it Muon
  Filter} before additional reconstruction could be performed. The
computing resources at the South Pole were expanded prior to the
second full season of IceCube in its 86-string configuration (IC-2012). 
This expansion made it possible to run some
reconstruction (a double-iteration SPE fit and the MPE fit) before
applying the {\it Online Level 2} cuts.

%The cuts of the Online Level 2 filter were designed such that this
%filter provides a very generic event sample, that could be used for
%different analyses. Therefore, the cuts chosen should be simple
%and stable. The selected events are a strict subset of the Muon Filter
%events. All reconstructions and cut variables applied before and after
%the Online Level 2 filter are stored and transferred via satellite to
%the North for all events that pass the filter.

In the up-going region, the main criteria to distinguish the
mis-reconstructed atmospheric-muon background from the neutrino events
are quality parameters of the reconstructed track. Several variables
derived from the single-iteration SPE fit were used to identify these
well-reconstructed tracks and to suppress mis-reconstructed air-shower
muons during the IC-2011 season. During  the IC-2012 season these
variables were derived from the MPE fit.  The most important variables
are:

\paragraph{The Scaled  Log-Likelihood}
\label{sec:ntoo:evtsel:onlineL2:up:rlogl}

In a maximum-likelihood fit the value of the likelihood at the maximum
divided by the number of degrees of freedom of the fit  can  measure 
the fit quality. The scaled likelihood of the track fit is
\begin{equation}
    S_{\textrm{LogL}}(x) = \frac{-\log_{10}(\max
      \mathcal{L})}{N_{\text{DOM}} - x}\,,
\end{equation}

where $N_{\text{DOM}}$ is the number of hit DOMs, and $x$ the numbers of parameters determined by the fit, usually five:  two angles, and three coordinates. However,  
it has been shown empirically that  $S_{\textrm{LogL}}(5)$ 
is energy-dependent for the track fits employed in IceCube. 
Thus, a cut  on   $S_{\textrm{LogL}}(5)$ variable introduces a bias
against well-reconstructed low-energy events. In order to reduce this
energy dependence bias, the  different values of $x$ is used i.e. $x=3.5$ for NToO and  $x=4.5$ for for optical and X-ray follow up.  This variable is especially useful for identifying mis-reconstructed muon tracks.

% For the SPE and
%MPE track fits
%employed in the Online Level 2 filter the reduced log-likelihood $S_{\text{LogL}}$ can be written as
%\begin{equation}
%    R_{\text{LogL}} =
%    \frac{-\log\mathcal{L}_{\text{max}}}{N_{\text{DOM}} - 5}\,
%\end{equation}
%where $N_{\text{DOM}}$ denotes the number of hit DOMs in the event.
%It has been shown empirically that this variable, however, is energy dependent
%for the track fits employed in IceCube. Thus a cut on this variable introduces
%a bias against well reconstructed low-energy events. In order to reduce the
%energy dependence a modification is used where the
%log-likelihood value is instead divided by $N_{\text{DOM}} - 2$: 
%\begin{equation}
%    P_{\text{LogL}} =
%    \frac{-\log\mathcal{L}_{\text{max}}}{N_{\text{DOM}} - 2}\,.
%\end{equation}
%In addition to a smaller energy dependence this cut variable also shows a higher signal
%efficiency at a fixed background passing rate compared to the traditional reduced log-likelihood.

\paragraph{Number of Direct Hits ($N_{\textrm{Dir}}$)}
\label{sec:ntoo:evtsel:onlineL2:up:ndirc}
Another measure of the track quality is the number of DOMs that have
registered a hit with a very small time residual $t_{\text{res}} \in
[{-15}{\text{ ns}},{75}{ \text{ ns}}]$ with respect to the
arrival time expected for Cherenkov emission from the reconstructed muon track. Such hits are called 
``direct hits``. The number of direct hits $N_{\text{Dir}}$ is calculated using only the
first registered photon in each DOM.\@ A photon causing a direct hit has undergone
less scattering in the ice and thus can contribute more information to the directional
reconstruction. The number of direct hits is therefore related to the quality of the track reconstruction.

\paragraph{Direct Length ($L_{\text{Dir}}$)}
\label{sec:ntoo:evtsel:onlineL2:up:ldirc}

% In a rotated coordinate system where the x-axis is given by the fitted track direction
% a quantity called ``Direct Length'' can be defined as
% \begin{equation}
%   L_{\text{Dir}} = |\max_{\text{Direct hits}} x - \min_{\text{Direct hits}} x|
% \end{equation}
% which is T

The distance between the projected direct hits onto the reconstructed
track is referred to as the ``direct length'', $L_{\text{Dir}}$. The
hits defining the direct length result from the least-scattered
photons and hence contribute the most to the reconstruction. If
$L_{\text{Dir}}$ is large, the lever arm for the reconstruction is
longer, generally resulting in smaller reconstruction
errors. Therefore, selecting events with larger $L_{\text{Dir}}$
selects the events most valuable for a point-source analysis.

%\par
%The Online Level2 filter selects events that were
%reconstructed as upgoing (zenith angle $\theta>82^\circ$, $\theta=0^\circ$ equals
%vertically down-going tracks) with a likelihood reconstruction that
%takes into account the time of arrival of the first photon at each
%Digital Optical Module (DOM) and the total charge recorded in that module.
%By requiring a good reconstruction quality the background of misreconstructed atmospheric
%muons is reduced. The parameters used to assess the track quality are
%the likelihood of the track reconstruction, the number of unscattered
%photons with a small time residual w.r.t.~the Cherenkov cone and the
%distribution of these photons along the track. 
The cut variables described above have been combined to achieve good
background rejection as well as reasonable signal efficiency. Events that
are reconstructed as up-going (zenith angle $\theta_{\text{SPE}} >
82^{\circ}$) and fulfill

\begin{eqnarray}
S_{\textrm{LogL}}(4.5) \leq 8.3 \; \text{or} \;   \text{log}_{10}(\frac{\text{QTot}}{p.e.})\geq 2.7 \;  \nonumber \\
 \text{or} \; \left( \frac{L_{\text{Dir}}[m]}{160}\right)^2 + \left( \frac{N_{\text{Dir}}}{9} \right)^2 \geq 1&
 \label{ellipse}
 \end{eqnarray}
where QTot is the total charge of the event in photoelectrons (p.e.), are selected by the {\it Online Level 2 Filter}. 
The  pre-selection criterion  based on the number of direct hits
($ N_{\text{Dir}} $) and the direct length ($ L_{\text{Dir}} $)  in  Eq.~(\ref{ellipse})  is called the ``direct ellipse`` cut.
The background  atmospheric-muon events tend to have short direct lengths
and a small number of direct hits since, if they are mis-reconstructed
as upward-going muon tracks, the hit pattern tends to match poorly.
The direct ellipse cut keeps $\sim 74$\% of atmospheric neutrinos and
$\sim 76$\% of astrophysical neutrinos while rejecting $\sim 93$\% of
atmospheric muons.

\paragraph{Quality  parameters of the  track reconstruction}
A critical parameter in a maximum-likelihood-based search for neutrino
point sources is the error of the reconstruction for each event.  As
this can only be determined on an event-by-event basis with simulated
data, an estimate has to be used for experimental data.  Two different
approaches are applied in IceCube: the { \it Paraboloid fit} and the
{\it Cram\'er-Rao Resolution Estimate}.  The {\it Paraboloid fit}
scans the likelihood space around the minimum determined in the track
fit by varying the fit parameters.  The resulting points in
likelihood space are fitted with a parabola~\cite{parabol}. Due to the
repeated evaluation of the likelihood function this method can be too
slow to be used in the online filtering, especially for high-energy
events with a large number of hit DOMs.  The {\it Cram\'er-Rao
  Resolution Estimate } is the uncertainty on the reconstructed track
direction given by the log-likelihood-based track reconstruction
estimated by a method based on the Cram\'er-Rao
inequality~\cite{cr}. As the calculation involves no minimization of a
likelihood it is considerably faster (and have a similar performance ) than the {\it Paraboloid fit} and
thus is the preferred method to be used in online analysis, also in NToO.  Since the
likelihood used in the track fit fully describes the Cherenkov
light emission and propagation, both angular uncertainty estimates,
the {\it Paraboloid fit} and the Cram\'er-Rao method, show an
energy-dependent bias in the ratios of the estimated to the true
angular uncertainty. This bias can be calibrated using Monte Carlo
events to derive a correction factor which is a function of the
reconstructed event energy.

The event rate after application of the {\it Online Level 2 Filter} is
reduced to approximately 2 Hz (in the up-going region). More than 99\%
of well-reconstructed signal neutrinos (i.e.\ reconstructed within
3$^\circ$ from their true direction) from an $E^{-2}$ energy spectrum
are retained, with respect to the number which pass the {\it Muon
  Filter}.

In subsequent analysis steps, more-computationally-intensive
reconstructions can be performed, including angular-resolution
estimators, energy estimators and likelihood fits applied to different
subsets of the recorded photons.

\subsection{NToO  selection variables}
The background rejection of the {\it Online Level 2 Filter} is still
not sufficient for NToO data analyses, since the sample is still
dominated by mis-reconstructed atmospheric muons. Only approximately 1
out of every 1000 events is a neutrino.  Thus, based on variables from
the {\it Online Level 2 Filter} reconstructions the final event sample
is selected by employing additional quality cuts. The following
additional cut variables are used:

%\paragraph{Fit Likelihood}
%
%The scaled likelihood of the track fit  $S_{\textrm{LogL}(x)}$.
%\begin{equation}
%    S_{\textrm{LogL}}(x) = \frac{-\log\mathcal{L}_{\text{max}}}{N_{\text{DOM}} - x}\,.
%\end{equation}
%where $N_{\text{DOM}}$ denotes the number of hit DOMs in the event and . In OnlineL2  $x=4.5$ for GFU/OFU and $x=3.5$ for NToO.  It has been shown empirically that this variable, however, 
%is energy dependent for the track fits employed in IceCube. Thus a cut on this variable introduces
%a bias against well reconstructed low-energy events. In order to reduce the
%energy dependence a modification is used where the log-likelihood value is instead divided by $N_{\text{DOM}} - 2$: 
%This variable is one of the most powerful cut variables to identify misreconstructed muon tracks
%and in general measure the fit quality.

\paragraph{Split Fits}

The track reconstruction for a correctly-reconstructed up-going track
should be stable against changes to the set of DOMs used for
reconstruction. On the other hand, for two coincident down-going muons
wrongly reconstructed as one up-going track, or for other cases of
mis-reconstruction, changes to the DOM set will have a much larger
impact on the reconstructed direction. This is the rationale for
splitting the DOM set used in the reconstruction into two parts and
subsequently performing a maximum-likelihood fit on each part
separately. Different criteria can be used to split the DOM set:
\begin{itemize}
    \item geometrical splitting divides the hits according to
their position with respect to the center of gravity (COG) of all hits. The
center of gravity is calculated as
\begin{equation}
\vec{x}_{\text{COG}} = \frac{\sum_{i=1}^{N_{\text{DOM}}} q_{i}  \times\vec{x}_i}{ \sum_{i=1}^{N_{ \text{DOM}}} q_{i} 	}
\end{equation}
where $\vec{x}_i$ are the positions of the individual hit DOMs and $q_{i}$ the charge of each hit DOM.
The center of gravity $\vec{x}_{\text{COG}}$ is then projected onto the track obtained
with the MPE fit, yielding the point $\vec{x}_{\text{COG}}^{\text{proj}}$\,. Each hit location is then also projected onto the track, and compared to $\vec{x}_{\text{COG}}^{\text{proj}}$\,.
Hits whose projections lie on one side of $\vec{x}_{\text{COG}}^{\text{proj}}$ are sorted into one set, hits whose projections lie on the other side are sorted into a second set.

\item time-based splitting divides the hits into two sets by comparing each
    hit to the mean of all hit times $t_{\text{mean}}$. Hits that fulfill $t_i \leq t_{\text{mean}}$ are sorted into one set, hits that fulfill $t_i > t_{\text{mean}}$ are sorted into another set. 

\end{itemize}

For each of the four subsets of hits a standard {\it Linefit}  is performed which acts as a seed
for a two-iteration SPE maximum-likelihood fit. The zenith angle $\theta_i$
resulting from the SPE fit, when only the time and charge of the first hit on each DOM are
taken into account in the reconstruction, is used to define the cut variable

\begin{equation}
    \Delta_{\text{Split/SPE}} = \max_{i \in \text{split
    fits}}(\cos{\theta_i} - \cos{\theta_{\text{SPE}}})\,.
\end{equation}

\paragraph{Bayesian Likelihood Ratio}

The probability that an event selected by the {\it Online Level 2
  Filter} and reconstructed as up-going (i.e.~zenith angle
$\theta_{\text{SPE}} > 90^{\circ}$) is truly a neutrino-induced muon
and not a mis-reconstructed air-shower muon is very small
($\sim10^{-3}$).  A useful additional cut variable can be derived by
forcing a down-going track fit and calculating the likelihood ratio to
the SPE fit.  This cut is motivated by a Bayesian approach
\cite{BAYES} to event reconstruction.  Bayes' Theorem in probability
theory states that for two assertions $A$ and $B$,
\[ P(A \, | \, B) \; P(B) = P(B \, | \, A) \; P(A), \]
where $P(A\,|\,B)$ is the probability of assertion $A$ given that $B$ is true. 
Identifying $A$ with a particular muon track hypothesis $\mu$,  and $B$ with
the data recorded for an event in the detector, we have
\[ P(\mu \, | \, \text{data} ) = 
\mathcal{L}_{\text{SPE}}( \text{data}\, | 
\, \mu) \;P(\mu),\]
where we have dropped a normalization factor $P(\text{data})$, which is a
constant for the observed event.  The function 
$\mathcal{L}_\text{SPE}$ is the regular SPE  likelihood function, and $P(\mu)$ is the
so-called prior function,  which is the probability of a muon  passing through the IceCube detector, and is  given by
\begin{equation}
     P(\mu)  = 2.50\cdot 10^{-7} \cos{\theta}^{1.68} \cdot
    \exp{\left(-\frac{0.78}{\cos{\theta}}\right)}\,,		
\end{equation}
This function is a fit  to the simulated zenith-angle distributions 
of  down-going cosmic ray muons  triggered by IceCube,  see also   \cite{BAYES2,bayesanfit} for  more details.
By accounting in the reconstruction for the fact that the flux of down-going
muons from cosmic-rays is many orders of magnitude larger than that of
up-going neutrino-induced muons, the number of down-going muons that are
mis-reconstructed as up-going is greatly reduced.

% In contrast to the SPE fit, where 
% only the time and charge of the first hit on each DOM is
% taken into account,  the MPE fit takes the time of the first hit, but the charge of all hits into account.
%The likelihood of this Bayesian fit is calculated as the product of the MPE
%likelihood of the track and a weight 
%\begin{equation}
%    w = 2.49655 \cdot 10^{-7} \cos{\theta}^{1.67721} \cdot
%    \exp{\left(-\frac{0.778393}{\cos{\theta}}\right)}\,.
%\end{equation}
%The weight $w$ is the  fit  to the  the muon air shower flux as a function of
%zenith angle~\cite{bayesanfit}. In order to force the numerical minimizer to reconstruct a down-going
%track ($0 \leq \theta \leq 90^{\circ}$) a penalty term is multiplied to the likelihood
%for track hypotheses with zenith angle  $\theta > 90^{\circ}$. 
\par
The difference of the logarithms of the SPE likelihood fit and the Bayesian-fit likelihood
\begin{equation}
\Delta_{\text{SPE/Bayesian}} =  \log_{10}{\mathcal{L}_{\text{SPE}}}- \log_{10}(\mathcal{L}_{\text{SPE}}( \text{data}\, | 
\, \mu) \;P(\mu) )
\end{equation}
is another useful cut variable, especially for events that have been reconstructed with a zenith angle close to the horizon.

%{\it {\bf Comment from MarkusV:} center of Gravity is charge weighted (I looked it up in the code,
%$I3_SRC/phys-services/private/phys-services/I3Cuts.cxx, l. 236-294)$, so
%the formula must contain the charge of each hit DOM,$ q_i$, multiplied
%with vector $x_i$ and the normalization is not $N_DOM$, but the sum over all
%charges.
%
%You say that an SPE fit is done on all split events, but you don't
%explain what SPE means. To be more precise, it should be called SPE1st
%(which is a technical term maybe not appropriate for a paper), and it
%means that only the time and charge of the first hit on each DOM is
%taken into account, whereas the MPE fit takes the time of the first hit,
%but the charge of all hits into account (as you explain on page 8, above).
%
%You use the MPE as subscript in eq. 4, but it's never defined. If you
%use it, then you should define where you explain MPE fit (Online L2
%part). You also don't clearly define theta as being the zenith angle.
%}

\subsection{NToO cut optimization}
\label{sec:ntoo:evtsel:neutrinosel:cutopt}

{
For neutrino searches, muons produced in cosmic-ray-induced air showers are the dominant background in IceCube for tracks coming
from the Southern Hemisphere. They trigger the detector with a rate $10^{6}$ times higher than atmospheric neutrinos.  Tracks from the Northern Hemisphere originate mostly from atmospheric neutrino interactions that produce muons. BOTH of these background components are simulated using Monte Carlo studies.

Cosmic-ray air showers are simulated using a patched version of  CORSIKA~\cite{corsika}.  The spectra of cosmic-ray primaries are sampled from the phenomenological {\it Polygonato} model \cite{polygonato} and the background datasets were produced with the Sybill~\cite{sibill} hadronic-interaction model.  A  sizable fraction of events in the IceCube detector include several muons from distinct air showers. These so-called coincident air-shower events are simulated as well.

Neutrino events are simulated based on  the Monte Carlo generator ANIS~\cite{anis}. ANIS generates neutrinos, propagates them through the Earth and finally forces them to interact in a volume around the detector. As different primary neutrino spectra are
needed by different analyses, one usually simulates a generic primary spectrum $dN/dE_{\mu}  \sim E_{\nu}^{-\gamma}$ where ${\gamma=1}$ or ${\gamma=2}$.
The events can be re-weighted to the desired spectrum for each analysis. The output of the
neutrino generator in the case of a charged-current
$\nu_\mu$ interaction is a muon produced at the interaction vertex and the accompanying hadronic cascade. The cascade is not simulated in detail.

The simulation of the muon propagation and the muon energy loss is essential to obtain the light distribution in the detector. A software package called
PROPOSAL~\cite{proposal} is employed for
that purpose. PROPOSAL calculates the continuous energy loss of the muon as well as the stochastic losses due to bremsstrahlung, pair production, photonuclear interactions and delta electrons. Finally, the detector simulation is concerned with the response of the PMTs to detected photons, the digitization of the PMT waveform in the DOM, and the trigger system. This  is done by the IceCube  software framework called  IceTray~\cite{icetray}.

}

In order to achieve the best possible sensitivity, the cuts on the variables described in the previous section have been optimized together in independent zenith-angle regions. For each combination of cut values, the rate of remaining data events was used as the approximation for the background rate. The rate of signal events for a given flux was estimated from neutrino simulations, assuming a neutrino flux with a spectral index $\gamma=2\,$. { This is motivated by the fact that  diffuse
shock acceleration leads to power-law spectra with a spectral
index around 2 \cite{Bell1978, Schlickeiser1989}  and neutrinos
originating in cosmic rays interactions near the source are expected to follow a similar spectrum. }
%
%\begin{table}[hbt]
%  \centering
%    \begin{tabular}{|c|c|c|}\hline
%  IceCube data taking period    &  Start  & End \\
%      \hline
%      IC-2011& 2011 May 13             & 2012 May 16 \\
%      IC-2012& 2012 May 16             & 2013 May 3 \\
%      IC-2013& 2013 May 3               & 2014 May 5 \\
%      IC-2014& 2014 May 5               & 2015 May 18 \\
%      IC-2015& 2015 May 18             & 2016 May 20 \\ 
%      
%      \hline
%
%    \end{tabular}
%    \caption[Neutrino selection cut efficiency]{Summary of  IceCube data taking periods (seasons) used by NToO searches }
%    \label{tab:ntoo:seasons}
%\end{table}

\begin{figure}[!htp]
  \centering
  \label{fig:ntoo:evtsel:neutrinosel:cutopt:10days}
    \includegraphics[width=0.6\textwidth]{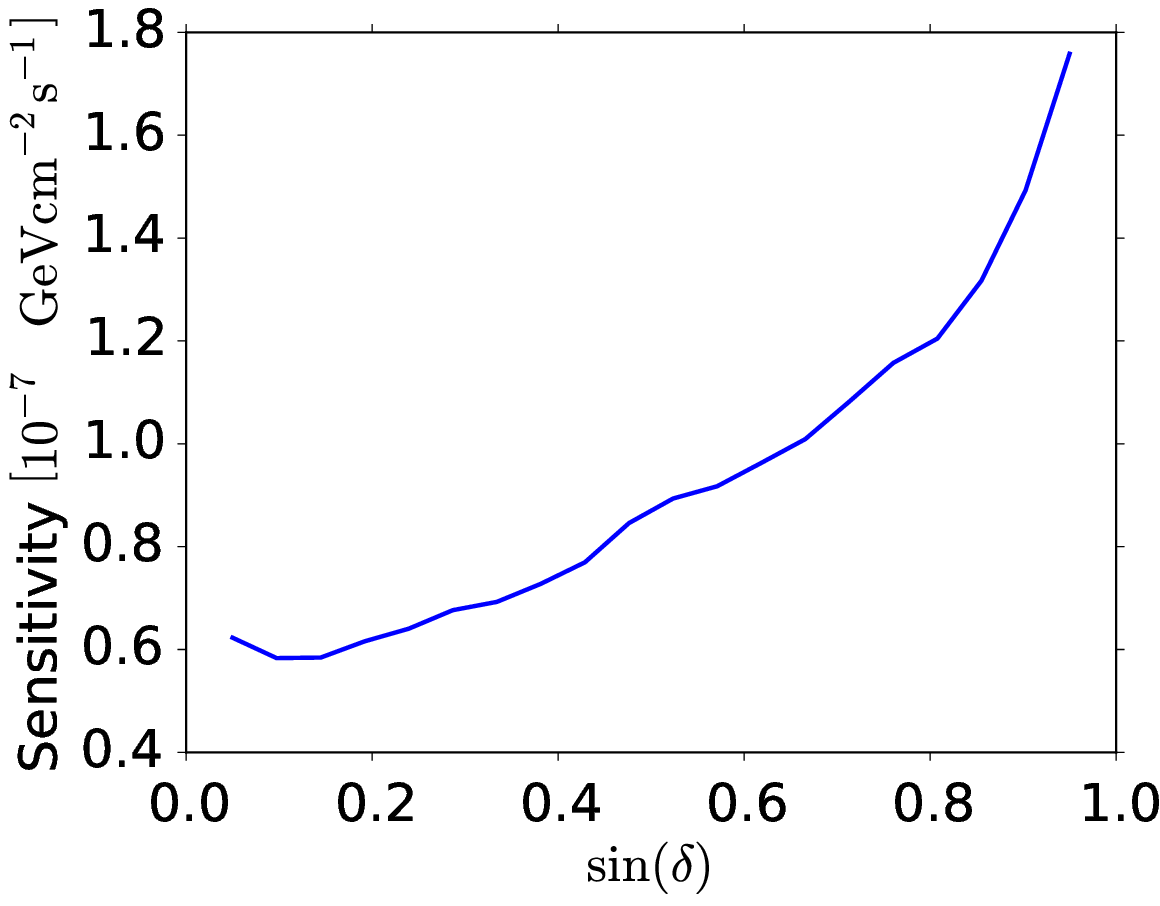}
\vspace{-0.4em}
\caption[IceCube sensitivity for different cut optimizations]{Simulated IceCube sensitivity as a function of source declination  to a neutrino flux with $dN/dE \sim E^{-2}$
for a neutrino flare length of  10 days and for the IC-2012 data set. The sensitivity did not change significantly for other IceCube  data seasons. The sensitivity  is defined as the flux required for a 3 sigma detection  with a probability of 90\%. }
\label{fig:ntoo:evtsel:neutrinosel:cutopt}
\end{figure}

The cut
optimization was repeated for flares of different durations ranging from $1$ day to $20$ days; as an example see
Figure~\ref{fig:ntoo:evtsel:neutrinosel:cutopt}. As traditional
minimizers like MINUIT~\citep{minuit} were found to converge on local minima a
particle  swarm  optimization algorithm  was used~\cite{KE95}. For
simplicity the minimization assumes that the flare time window is
known.

In the binned point-source method the radius of the on-source bin is a
free parameter. The optimal bin size as a function of zenith depends
on the angular resolution and the background rate of atmospheric
neutrinos at each zenith angle. The search-bin radius has also been
optimized together with the cut variables to yield the best
sensitivity.

{ The first  set of the NToO  cuts was optimized  using the IC-2011  season, and later redefined using 
data from the IC-2012  season. Only the cut on the Bayesian likelihood ratio was changed.}

\begin{table*}[htb]
  \centering
 \small
  \begin{tabular}{|c||c|c|c|c|c|}\hline
    Cut Level  & Selection criterion& Atms.  $\mu$& Data& Atms. $\nu_{\mu}$ & Astro.\\
                     &  & (mHz)& (mHz)& (mHz) & $ \times 10^{-3}$ (mHz)\\
      \hline
      0& $ \cos{\theta_{\text{MPE}}} \leq 0 $ & 1010.5 & 1523.81 & 7.166& 6.23 \\
      1 &  \text{SLogL(3.5)}  $\leq 8$ & 282.49&504.44 & 5.826& 5.62 \\
      2 & $N_{\text{Dir}} \geq 9$ & 8.839&  22.01& 3.076 &4.06 \\
      3 &   ($ (\cos{\theta_{\text{MPE}}} > -0.2)  \text{ AND } (L_{\text{Dir}} \geq 300 \text{ m})$   & &  &  & \\

 &  OR  & 1.124 & 4.30 &2.313 & 3.69 \\
       &   $ (\cos{\theta_{\text{MPE}}} \leq -0.2)  \text{ AND } (L_{\text{Dir}} \geq 200 \text{ m})$) & &  && \\
 
   4 &   $\mathbf{\Delta_{\text{Split/MPE}}}$  <0.5 & 0.100&  2.15& 1.899 & 3.26\\
 %    &&&&\\
   5  &   ($ (\cos{\theta_{\text{MPE}}} \leq -0.07)  $ & &  && \\
   
       &   \text{ OR }    & 0.080& 2.08 &1.880&3.25\\
       &   $  (    ( \cos{\theta_{\text{MPE}}} > -0.07)   \text{ AND }  (\mathbf{\Delta_{\text{SPE/Bayesian}}}   \geq 35  ) )$ )  & &  && \\
   
%   &&&&\\
    6  & (  $( \cos{\theta_{\text{MPE}}} \leq -0.04)  $   & & &&\\         
    
    &    \text{ OR }    & 0.075& 2.06 &1.875 &3.24\\     

    &   $ (    ( \cos{\theta_{\text{MPE}}} > -0.04)   \text{ AND }  (\mathbf{\Delta_{\text{SPE/Bayesian}}}   \geq 40) )$) & & &&\\      
      \hline
    \end{tabular}
    \caption[IceCube neutrino selection cuts 2012/2013]{IceCube neutrino selection cuts and corresponding passing event rate for the IC-2012 season. At an final selection an event has to fulfill all cut criteria to pass the selection (i.e. a logical AND condition between the cut levels is applied). The atmospheric-neutrino flux is based on the prediction by Honda \cite{honda}, but  atmospheric-muon rate is calculated  from   CORSIKA simulations. The event rate  for IceCube data stream corresponds   
   to the  total livetime  of 332.36 days. The astrophysical neutrino flux is estimated  assuming $ dN/dE = 1 \cdot 10^{-8} \text{ GeV} \text{cm}^{-2} \text{s}^{-1} (\frac{E}{\text{ GeV} })^{-2}$.    
  (Atms. = atmospheric, Astro. = astrophysical)}
  \label{tab:ntoo:evtsel:cutopt:cuts_2012}
\end{table*}

The final set of smooth cuts
resulting from the cut optimization is listed in Table~\ref{tab:ntoo:evtsel:cutopt:cuts_2012}
%\begin{table*}[htb]
%  \centering
% \small
%  \begin{tabular}{|c||c|c|c|c|c|}\hline
%      Region & $N_{\textrm{Dir}}$& \textbf{SLogL(3.5)} & $L_{\text{Dir}}$& $\mathbf{\Delta_{\text{Split/MPE}}}$ & $\mathbf{\Delta_{\text{SPE/Bayesian}}}$\\
%      \hline
%      $\cos{\theta_{\text{MPE}}}<-0.2$ & $\geq 9$ & $\leq 8.0$ & $\geq 200$ & $<0.5$ & No cut \\
%      $-0.2 \leq \cos{\theta_{\text{MPE}}} < -0.07 $ & $\geq 9$ & $\leq 8.0$ & $\geq 300$ & $<0.5$ & No cut \\
%      $-0.07 \leq \cos{\theta_{\text{MPE}}} < -0.04 $ & $\geq 9$ & $\leq 8.0$ & $\geq 300$ & $<0.5$ & $>35$ \\
%      $-0.04 \leq \cos{\theta_{\text{MPE}}} \leq 0 $ & $\geq 9$ & $\leq 8.0$ & $\geq 300$ & $<0.5$ & $>40$ \\
%      \hline
%    \end{tabular}
%    \caption[IceCube neutrino selection cuts 2012/2013]{IceCube neutrino selection cuts for the $2012/2013$ season. An event has to fulfill all cut criteria to pass the selection (i.e. a logical AND condition between the cut variables is applied).}
%    \label{tab:ntoo:evtsel:cutopt:cuts_2012}
%\end{table*}
and the optimal search-bin radius as a function of declination angle ($\delta >0^{\circ}$)  has been parametrized as
\begin{equation}
r = 1.2^\circ + 1.4^\circ \cdot \sin(\delta)\,.
\end{equation}

{ Table~\ref{tab:ntoo:evtsel:cutopt:cuts_2012}  shows also  the  influence of  each selection  cut on event rate  for  data, simulated atmospheric neutrinos and muons.
 
The experimental data sample, after application of the {\it Online Level 2 Filter} (Cut Level 0), consists of $4.3 \times 10^{7}$ events acquired within a total livetime of 332.36 days. At this level,  
see  Table~\ref{tab:ntoo:evtsel:cutopt:cuts_2012},  atmospheric muons   dominant the contribution to the  measured data sample. However, these  mis-reconstructed events being truly down-going and reconstructed as up-going are  almost removed by our selection criterion, i.e.  the passing rate is reduced by  99.9925 \% with respect  to Cut Level 0.
We also see that at  the final selection cut   the data rate  reach the level of atmospheric muon neutrinos, i.e. about 2 mHz, and selection criterion
keeps  about  52\% of the  signal events (with respect to Cut Level 0)  for an $E^{-2}$ signal neutrino spectrum of astrophysical neutrinos.

The same  set of cuts was  used for the next IceCube seasons: IC-2013 and IC-2014 thanks the very stable behavior of the  IceCube detector and its performance. }

\subsection{Properties of the neutrino sample}

\begin{figure}[!htbp]
  \centering
    \subfigure[tight][] {
\label{fig:final_2012_angular_resolution:energy}
    \includegraphics[width=0.48\textwidth]{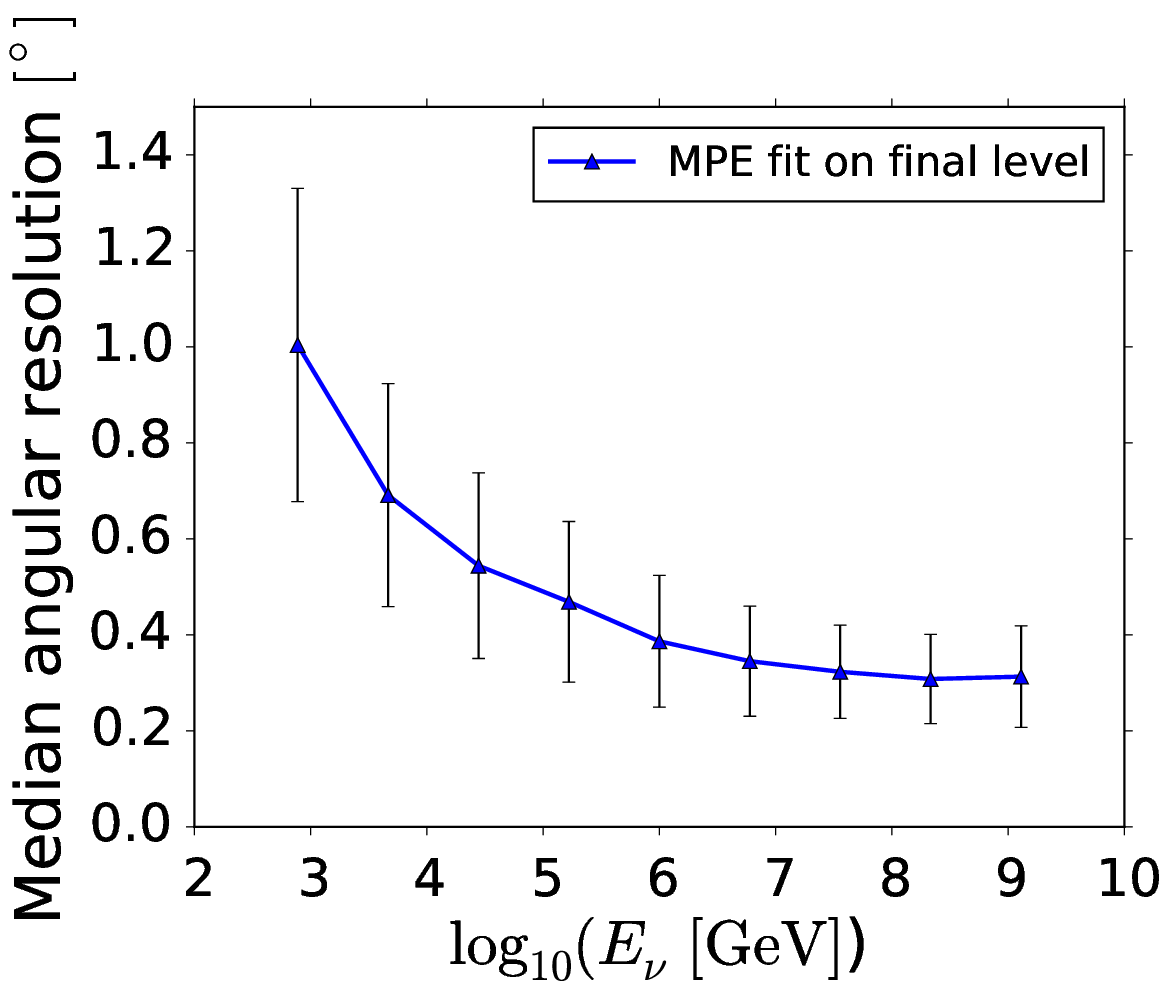}}
    \subfigure[tight][] {
\label{fig:final_2012_angular_resolution:zenith}
    \includegraphics[width=0.48 \textwidth]{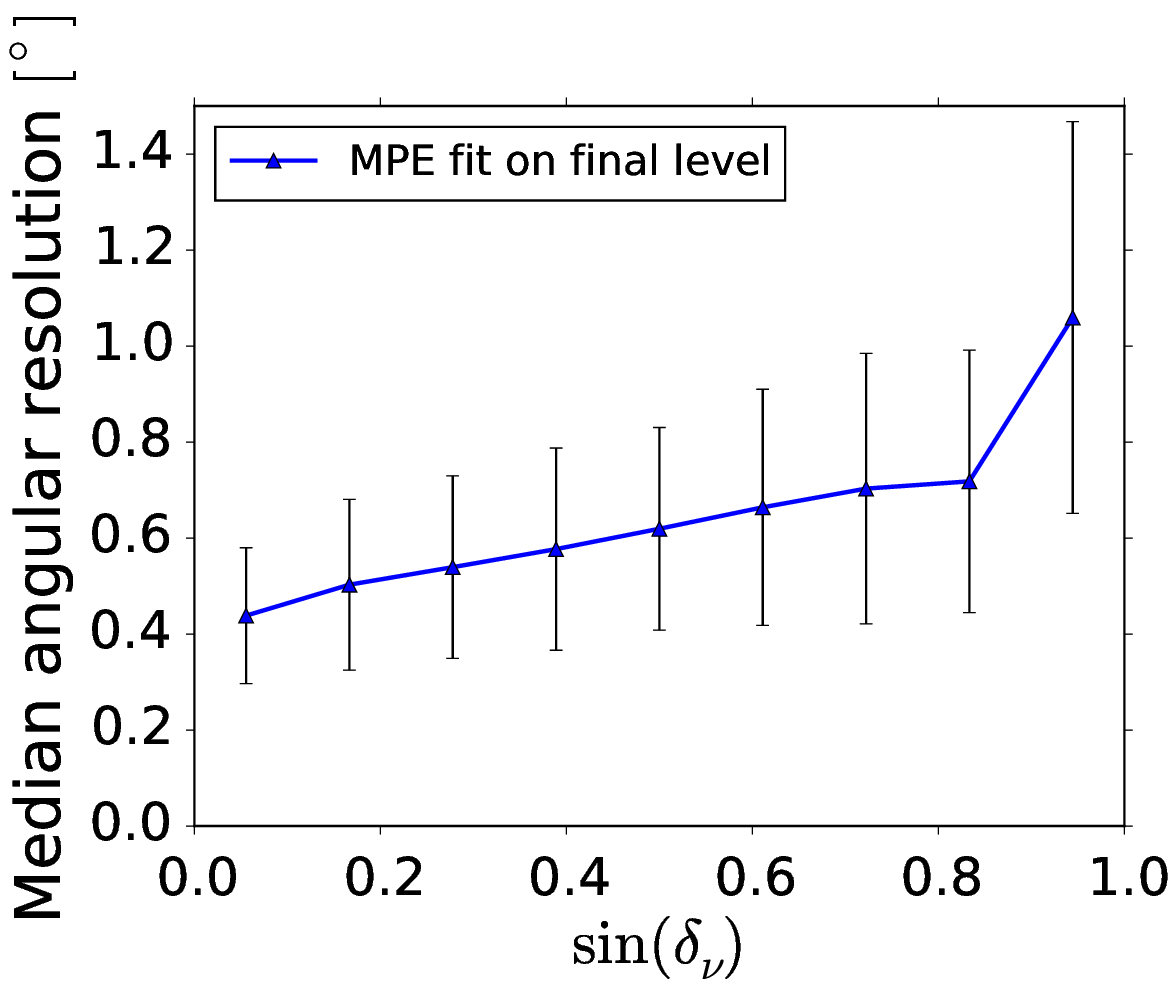}}
\vspace{-0.4em}
\caption[IceCube angular resolution of the $2012/2013$ neutrino sample]{IceCube median angular resolutions  based on { \it Cram\'er-Rao  Resolution Estimator }  for the final selected neutrino sample as
    a function of  (true) neutrino energy (left panel) and declination
    (right panel), assuming in simulation a primary neutrino spectrum with $\Phi(E) \sim E^{-2}$). The error bars depict the 16th and 84th percentile of 
the angular resolution. Neutrino angular resolution defined as the median of the point-spread function of the true neutrino direction and the reconstructed muon direction.} 
\label{fig:final_2012_angular_resolution}
\end{figure}

The event selection results in an event rate of about  $2\,$mHz and a median
angular resolution of $0.5\,^\circ$ for an $E^{-2}$ signal neutrino spectrum.
Figure~\ref{fig:final_2012_angular_resolution} depicts the median angular resolution of
the final neutrino sample as a function of neutrino energy and as a function
of neutrino declination angle~\footnote{IceCube is located at the South Pole, so the relation between zenith angle  and declination is given by simply transformation: $\theta=\delta+90^{\circ}$. }.

\begin{figure}[!htbp]
  \centering
    \includegraphics[width=0.6\textwidth]{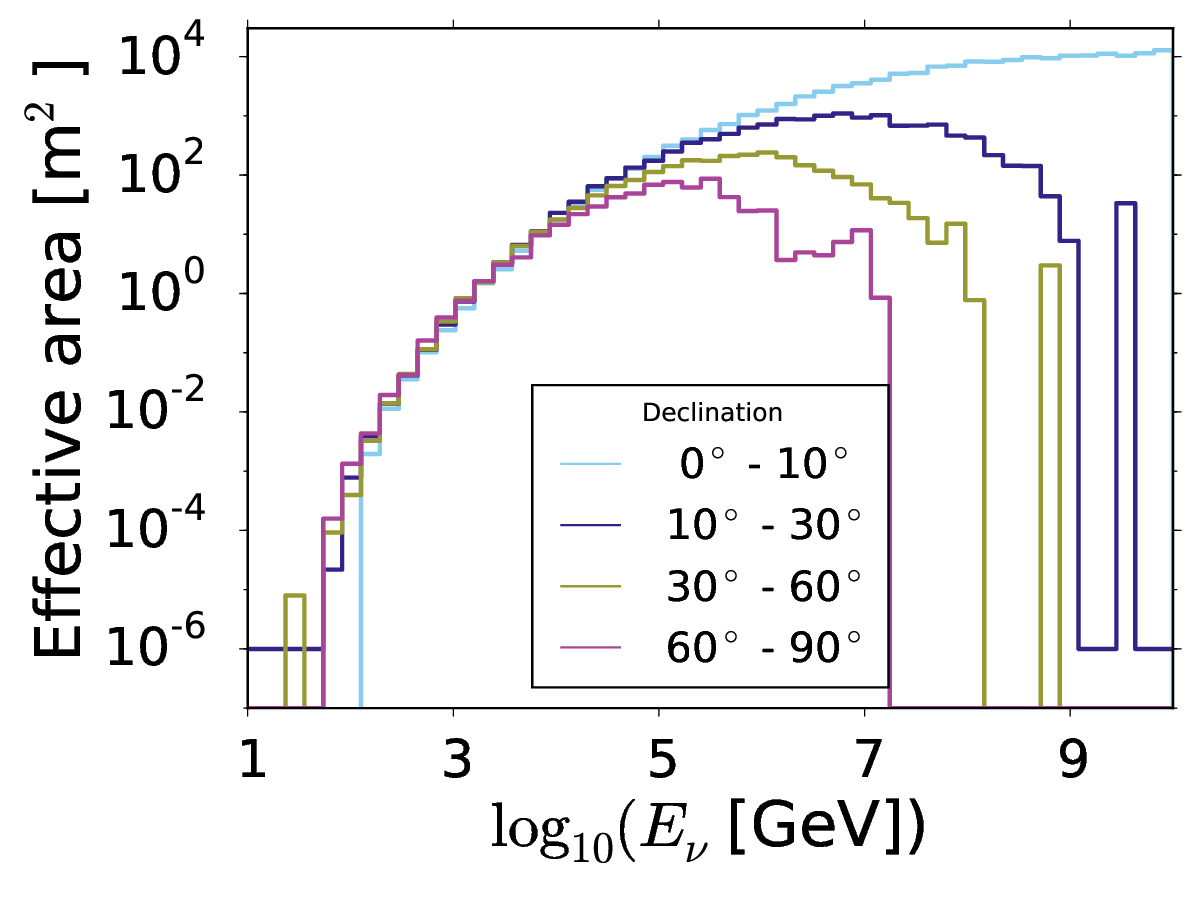}
\caption[Effective area on the final cut level]{Simulated IceCube effective area as a function
of the (true) neutrino energy for the 
  final neutrino selection derived from the  $2012/2013$ data set. The strongly
  energy dependent neutrino-nucleon cross section leads to the
  observed behavior of an effective area that is generally increasing
  with energy, until neutrino absorption dominates. For larger
  declinations the effect of neutrino absorption sets in at lower
  energies due to the longer path through the Earth. The  acceptance is similar  for other IceCube  data seasons. }
\label{fig:final_2012_Aeff}
\end{figure}

Figure~\ref{fig:final_2012_Aeff}  depicts the effective area for muon neutrinos
as a function of neutrino energy in different declination regions.
It is worthwhile to note that the effective area reaches only about 1
m$^2$ at $10^{3.2}$ GeV. For events with declination between
$10^{\circ}$ and $30^{\circ}$ the effective area reaches a maximum of
about 1000 m$^2$ at $10^{6.5}$ GeV and begins to drop above $10^{7.5}$
GeV due to absorption of neutrinos in the Earth. For neutrinos very
close to the horizon ($0^{\circ} \leq \delta \leq 10^{\circ}$) and for
neutrino energies greater than $10^{8}$ GeV the effective area can
reach $10^{4}$ m$^2$.
\par
\begin{figure}[!htbp]
  \centering
    \includegraphics[width=0.5\textwidth]{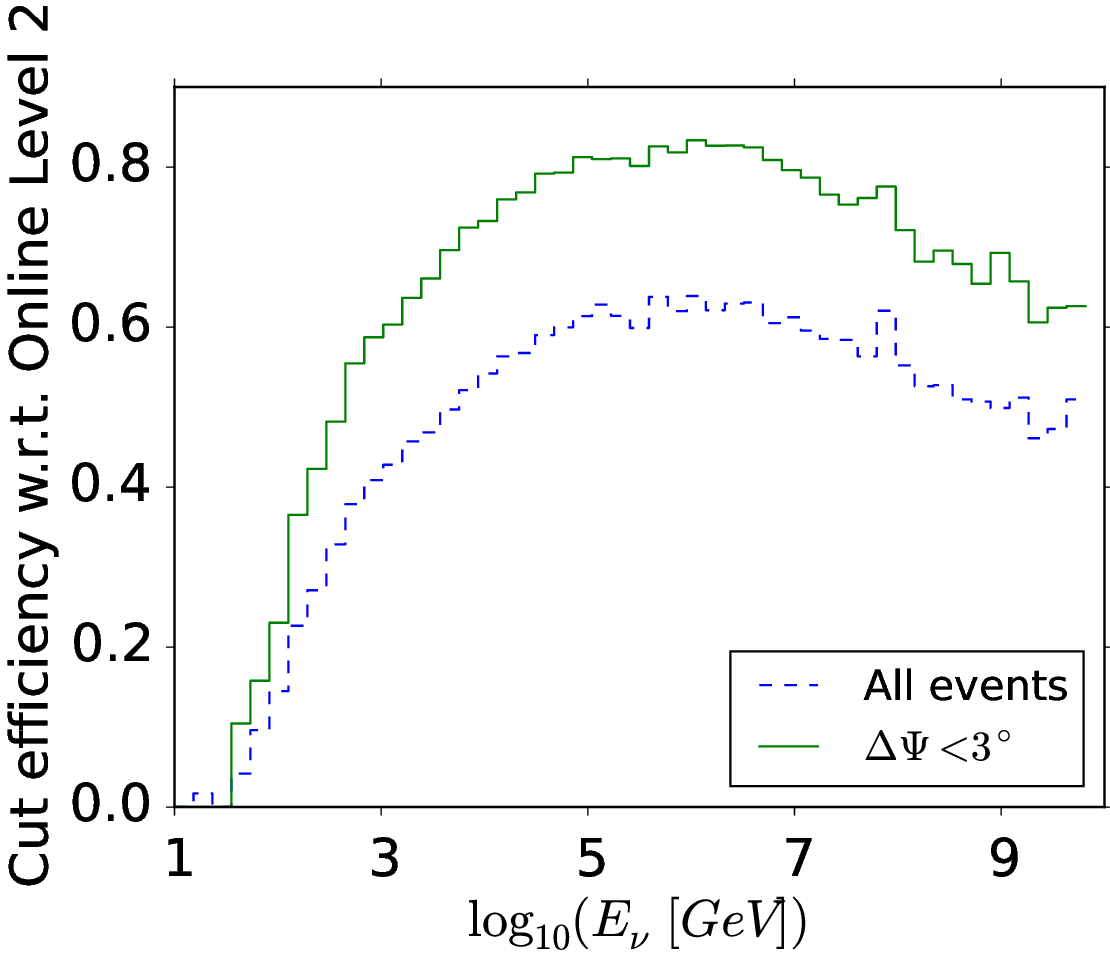}
\caption[Cut efficiency w.r.t.  Online Level 2]{Efficiency of the neutrino-selection cuts for IceCube  with respect to {\it Online Level 2 }  for all events (blue, dashed
line) and events with angular reconstruction error, $\Delta\Psi<3^{\circ}$ (green, solid line).}
\label{fig:final_2012_efficiency}
\end{figure}

 The efficiency of the event-selection cuts
with respect to the {\it Online Level 2 Filter} is depicted in
Figure~\ref{fig:final_2012_efficiency} for all events (dashed) and
for events that have been reconstructed within  angle $ \Delta\Psi<3^\circ$ of their true
direction. Well-reconstructed events are selected with an efficiency
of more than $60\,\%$ above 1 TeV; while the overall peak efficiency of
about $80\,\%$ is reached between 100 TeV and 10 PeV.

{As we already mentioned  above,  the main selection 
cuts are optimized for a neutrino power-law spectrum with a spectral index $\gamma=2\,$.
However,  several Galactic gamma-ray sources have energy spectra with energy cutoffs at a few TeV \cite{Mandelartz}, supporting the idea that Galactic neutrino
spectra may present cutoff spectra as well \cite{Kistler,Vissani}. Also, recent IceCube results show that the astrophysical neutrino flux  has a  neutrino spectrum softer than $E^{-2}$.
The IceCube neutrino flux can be  well described by an unbroken power law with best-fit spectral index $2.50\pm 0.09$~\cite{icecubespectrum}.

Therefore  in Table~\ref{tab:ntoo:evtsel:final:2012} the efficiency
of neutrino selection for  softer spectral indexes (e.g. 2.5 and 2.7) is also shown. As we can see, for softer spectra, the performance  of NToO cuts is  about 20 \% worse, but  the signal efficiency is still above 50\% for well-reconstructed events.
  }
\begin{table}[hbt]
  \centering
    \begin{tabular}{|c|c|c|}\hline
      $E^{-2}$ ($\Delta\Psi_{\text{MPE}}<3^{\circ}$)& $E^{-2.5}$ ($\Delta\Psi_{\text{MPE}}<3^{\circ}$) & $E^{-2.7}$ ($\Delta\Psi_{\text{MPE}}<3^{\circ}$)\\
      \hline

      $52\,\%$ ($73\,\%$)         & $43\,\%$ ($63\,\%$)               & $39\,\%$ $(59\,\%)$ \\
      \hline

    \end{tabular}
    \caption[Neutrino selection cut efficiency]{Efficiency (from simulation) of the neutrino selection cuts with respect to the {\it Online Level 2 Filter} (in \%).
      The efficiencies for well-reconstructed events (defined as events with $\Delta\Psi_{\text{MPE}}<3^{\circ}$) are given in parentheses.}
    \label{tab:ntoo:evtsel:final:2012}
\end{table}

\section{The time-clustering algorithm}
\label{time_clustering_algorithm}

The timescale of a neutrino flare is not fixed {\it a priori} and thus a simple
rolling-time-window approach is not sufficient to detect flares. The time-clustering approach that was developed for an unbiased neutrino flare search
~\cite{timeclustering} looks for any time frame with a significant deviation of the
number of detected neutrinos from the expected background. The simplest
implementation uses a spatially binned approach where neutrino candidates within a fixed
radius around a source are regarded as possible signal events.

\begin{figure}[!htbp]                                                                                                                                                       
  \centering                                                                                                                                                                  
  \includegraphics[width=0.6\textwidth]{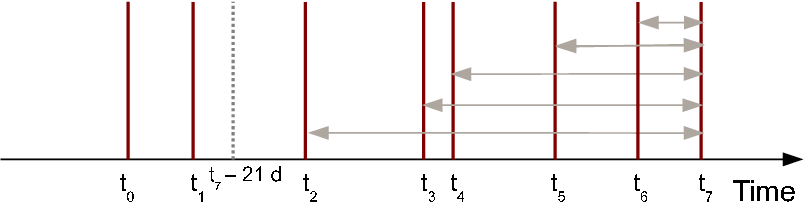}                                                                                                       
  \caption[Schematic of the time-clustering algorithm]{Schematic of the                                                                                                     
  time-clustering algorithm. For an IceCube event in an on-source bin detected 
  at time $t_7$ the  significances of all clusters formed with events detected up to $21$ days back                                                                                              
  are calculated.}                                                                                                                                                            
  \label{fig:ntoo:time_clustering}                                                                                                                                            
\end{figure}     

%  To exploit the
% information that can be extracted from the estimated reconstruction error and
% other event properties like the energy an unbinned maximum-likelihood
% method is under development.

\begin{figure}[!htbp]
  \centering
    \includegraphics[width=0.6\textwidth]{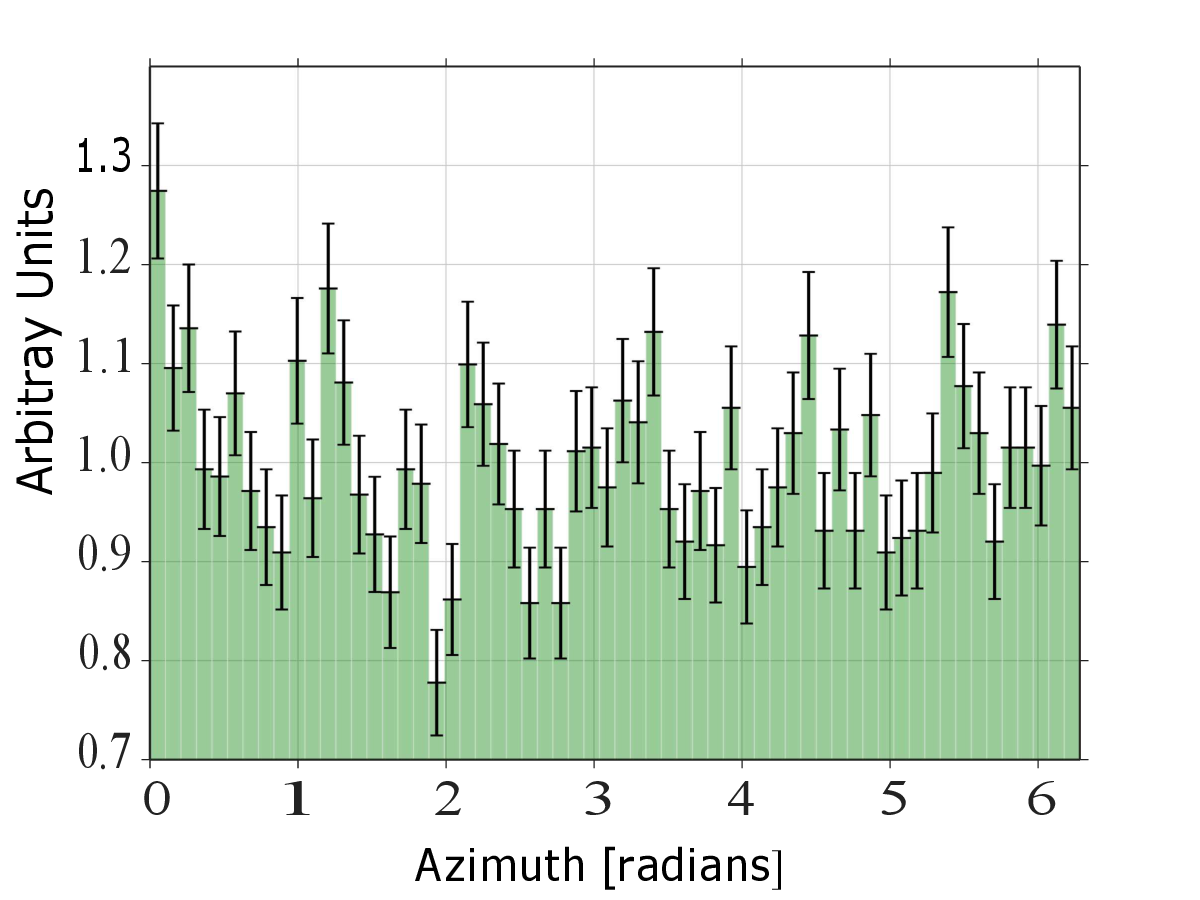}
\vspace{-0.4em}
\caption[Azimuth efficiency asymmetry]{The normalized  distribution of IceCube events  as a function of  azimuth. The dependence is caused by the hexagonal layout of the grid of IceCube
strings that produces symmetry axes.}
\label{fig:neutrinosel:azi_asymm}
\end{figure}

If a neutrino event  is detected at time $t_i$ from the search bin around a given source,
the expected background $N_{\tiny{\mbox{bg}}}^{i,j}$ is calculated for all other
events $j$ within a time window $\Delta t_j=t_j- t_i$ around that bin (see Figure~\ref{fig:ntoo:time_clustering}). To calculate 
$N_{\tiny{\mbox{bg	}}}^{i,j}$ the detector efficiency as a function of
the azimuth angle and the uptime has to be taken into account. 
The number of expected background events $N_\text{bg}^{i,j}$ in a time window
$[t_i, t_j]$ for a source at a certain declination is given by
\begin{equation}
    N_{\text{bg}}^{i,j}= t_{\text{up}}^{i,j} \dot{N}(\theta)
\epsilon(\Phi(t))
\end{equation}
where $\dot{N}(\theta)$ is the zenith-angle-dependent rate of  data  used as background events, $t_{up}^{i,j}$  uptime 
in a time window $[t_i,t_j]$ and $\epsilon(\Phi(t))$ the  normalized azimuth  distribution
of  IceCube  events (see Figure~\ref{fig:neutrinosel:azi_asymm}).

The Poisson probability of observing the multiplet $(i,j)$ by chance is then calculated
according to
\begin{equation}
    p_{\text{obs}} = \sum_{k=N_{\text{obs}}^{i,j}-1}^{\infty}
    \frac{(N_{\text{bg}}^{i,j})^k}{k!}e^{-N_{\text{bg}}^{i,j}}
    \label{eq:ntoo:nobs}
\end{equation}
\noindent
where $N_{\text{obs}}$ is the number of detected on-source neutrinos between
$t_j$ and $t_i$. $N_{\text{obs}}$  must be reduced by one to take into account the bias
introduced by the fact that the background is measured in the time window 
defined by the $j^{th}$ event.
Most  very high energy  flares detected so far  have a duration up to several days, thus 
we constrain our search for time clusters of neutrinos
to $21$ days so as to minimize the statistical trial factor penalty.

The probability $p_{\text{obs}}$ is often expressed in terms of the
distance to the center of a normal distribution measured in units of
standard deviations that results in the same cumulative probability in
the right tail (e.g.~a probability of
$\log_{10}{(p_{\text{obs}})}=-2.87$ is often quoted as $3\,\sigma$). If
the cluster with the highest significance exceeds a certain threshold
(e.g.~corresponding to $3\,\sigma$) the detector stability is first
checked and, if appropriate, an alert is sent to a partner experiment
(atmospheric-Cherenkov telescope) to initiate a follow-up observation.

The physical layout of the IceCube, with the instrumented strings
positioned on a hexagonal grid, results in an increased trigger rate  for events that propagate along the symmetry axes.
Therefore, the expected number of background events in a time window for a source at a certain
right ascension depends on the azimuth-angle range covered during that time.
This natural azimuth dependence is reinforced by cut variables that favour
events that pass close to many strings (e.g.~direct hits and direct
length).
For time-integrated point-source searches, the azimuth dependence is usually 
neglected because it is smoothed in right ascension by the rotation
of the Earth over long integration times.
However, in a time-dependent analysis the azimuth dependence becomes important for timescales shorter than $t_i -t_j < 2\,$ days.

The stable uptime between $t_{up}^{i,j}$ in a time window $[t_i,
t_j]$ is calculated using the online detector stability monitoring (described
in Sec.~\ref{data_stability_monitoring}) and combined with information about the start and
stop times of the data-taking runs.
                            
\subsection{Alert rate, detection probability}

Since the NToO aims at the discovery of neutrino flares from
astrophysical sources, it is important to define what astrophysical
neutrino flux level would trigger an alert.

\begin{figure}[!htbp]
\centering
\includegraphics[width=0.48\textwidth]{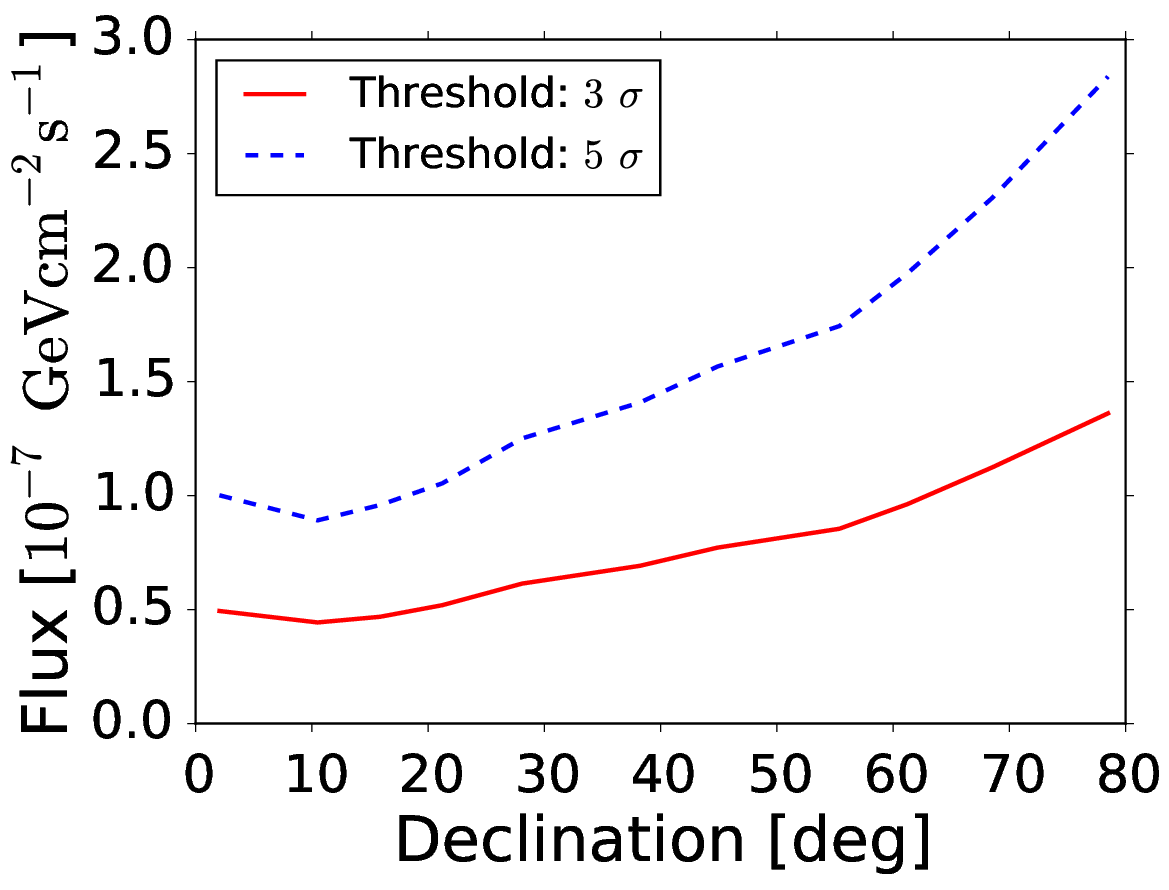}
\includegraphics[width=0.48\textwidth]{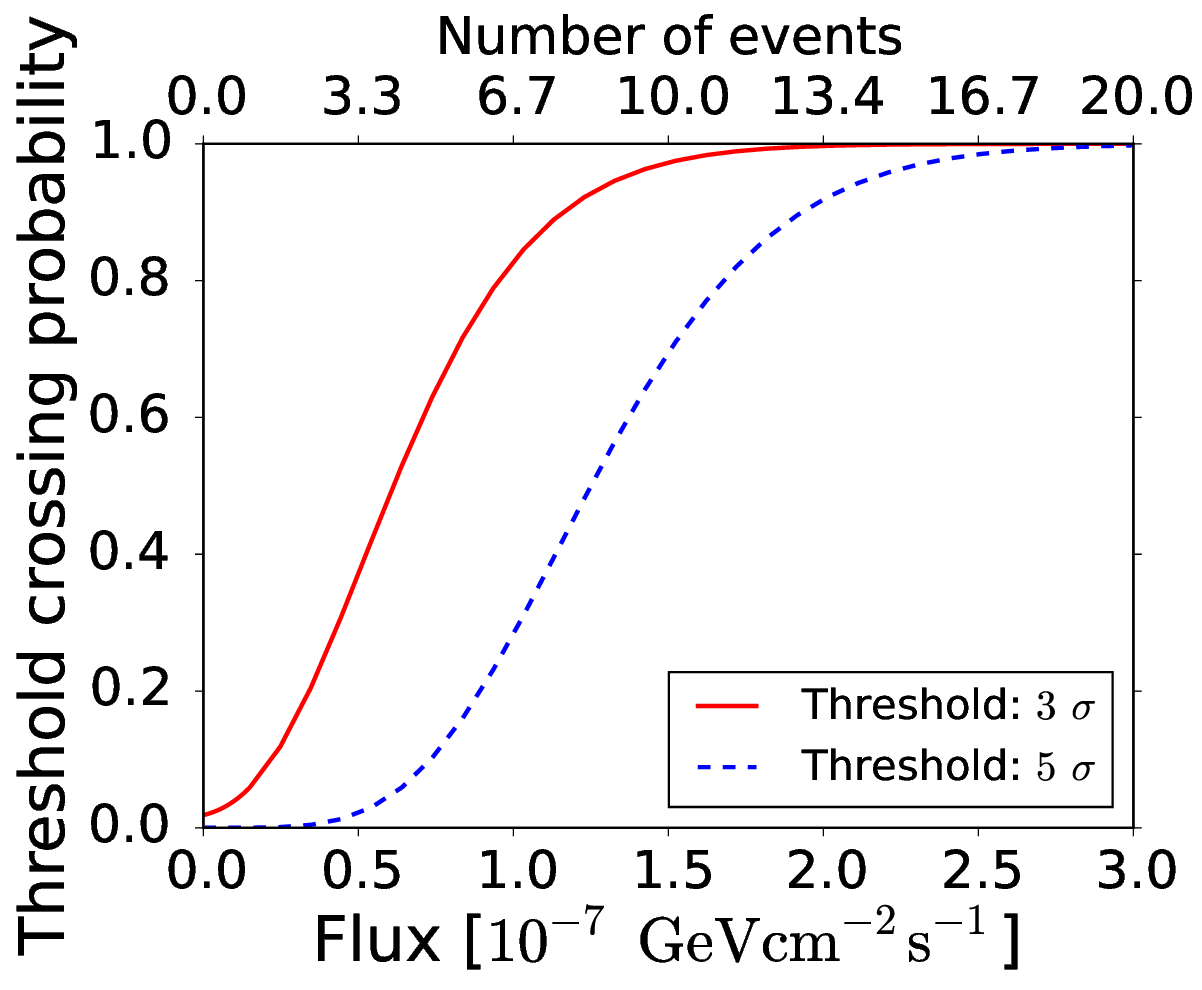}
\caption[IceCube alert trigger probability at $\delta=49.5^{\circ}$ as a function of flux]{(Left panel) 
Neutrino flux required by IceCube at a given source declination to trigger an alert with a significance of  $3\,\sigma$ (solid line) and   $5\,\sigma$ (dashed line) with a probability of 50\%.The neutrino spectrum  is assumed to be an unbroken power law
with a spectral index of 2; the flare duration is 10 days. (Right panel) Probability to trigger an NToO alert as a function of flux for flares with a duration of $10$ days at a declination $\delta=28.0^{\circ}$, for alert thresholds of $3\,\sigma$ (red, solid line) and $5\,\sigma$ (blue, dashed line). The upper axis shows the number of required events  needed for neutrino flux for alert with given significance. }
\label{fig:ntoo:alert:trig_prob_4c1554}
\end{figure}

In Figure~\ref{fig:ntoo:alert:trig_prob_4c1554} (A) we show the flux as a function of declination
that results in a trigger probability of 50\% for significance thresholds corresponding to $3.0\,\sigma$
and $5.0\,\sigma$ for a flare duration of 10 days, while in Figure~\ref{fig:ntoo:alert:trig_prob_4c1554} (B) the probability of triggering an alert as a function of the neutrino
flux, assuming a spectral index $\gamma=2$ and a flare duration of $10$ days,
is shown for alert thresholds corresponding to $3\,\sigma$ and $5\,\sigma$ for a source at a declination $\delta=28.0^{\circ}$.
As we can see from these plots, a signal flux of the form $\Phi(E) = 6 \cdot 10^{-8} \text{ GeV} \text{cm}^{-2} \text{s}^{-1} (\frac{E}{\text{ GeV} })^{-2}$ will on average trigger an
alert with a probability of 50\% for an alert threshold of $3.0\,\sigma$.

\begin{figure}[!htbp]
\centering
  \includegraphics[width=0.6\textwidth]{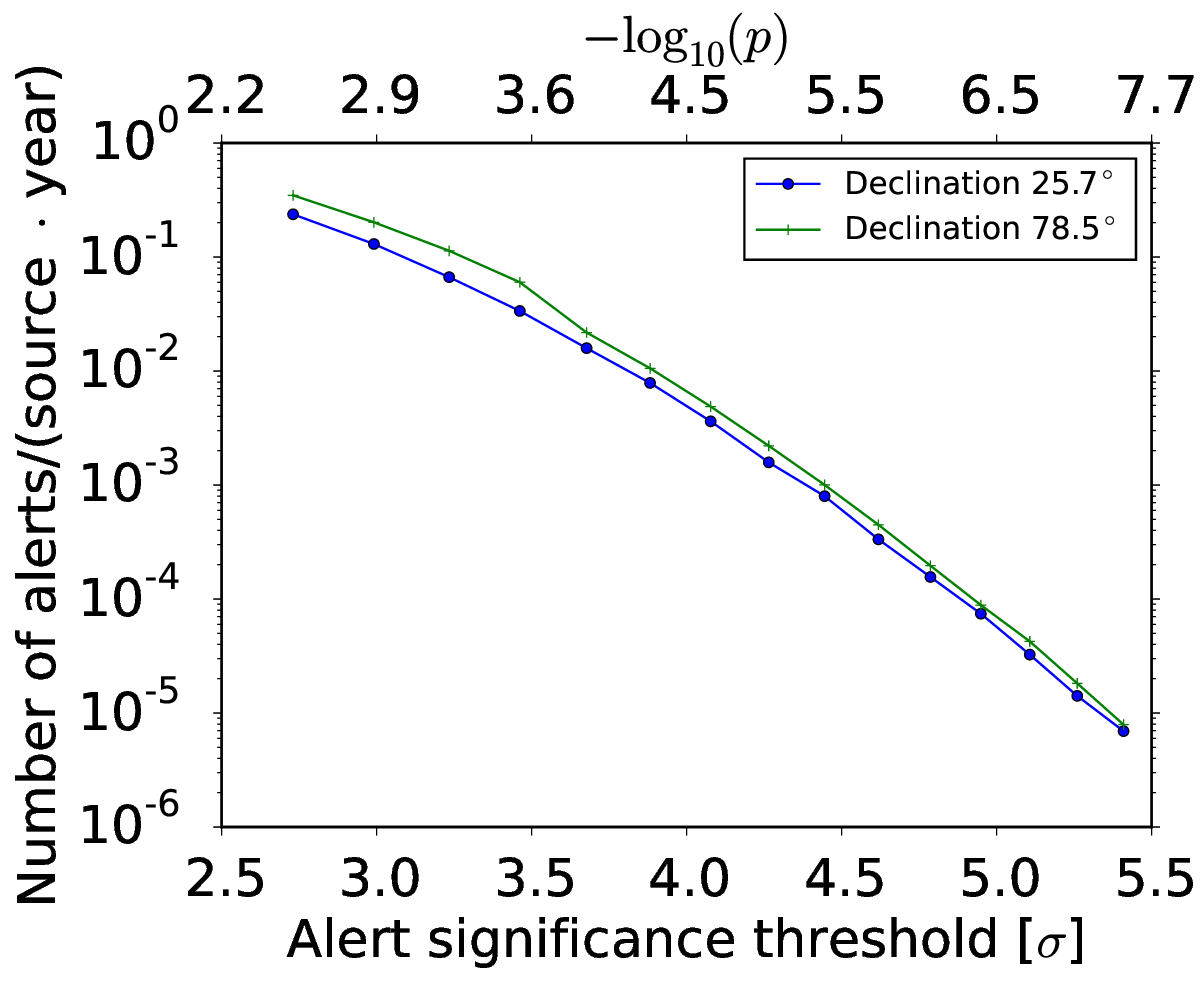}
\caption[Expected IceCube alert rates from atmospheric neutrinos]{Expected IceCube fake alert rates for the NToO caused by atmospheric neutrinos for different source
    declinations as a function of alert significance.}
\label{fig:ntoo:alert:bgalert}
\end{figure}

The number of accidental background alerts also needs to be estimated
in order to calculate the total significance of all the alerts generated
by the program, as well as to set sensible alert thresholds such that
the partner observatory is not overwhelmed by follow-up
requests. The number of follow-up requests allowed in a given time
period is fixed by the Time Allocation Committees of the partner
experiments.  Figure~\ref{fig:ntoo:alert:bgalert} shows the number of
accidental background alerts as a function of the alert significance
threshold.  For a threshold of $3.2\,\sigma$ (MAGIC) this would result
in a fake alert rate of about $0.1\mbox{ alerts}/(\mbox{source} \cdot
\mbox{year})$. Thus, given the number of sources (around 70) in this
program for the MAGIC experiment this results in about 3 background
alerts per year \footnote{Since November 2013 the number of MAGIC
  sources was lowered to 18 and since April 2014 the alert threshold was
  set to $3.6\,\sigma$, therefore the expected number of alerts
  decreased to one alert per two years.} . This number is calculated taking into account that
a source is on average visible with a probability of $40\%$ by a
partner observatory (i.e. if the source rises at least 30 degrees above
the horizon for at least 30 minutes during dark time, the current
phase of the Moon is less than $0.5$ and the source distance to the
Moon is larger than $60^{\circ}$).
For  VERITAS, a higher alert threshold ($3.6\,\sigma$)
leads to one expected background alert per year.

% Figure~\ref{fig:final_2012_sensitivity} shows the sensitivity of the final
% sample for neutrino flares with a spectrum $\Phi(E) \sim E^{-2}$ and flare
% durations of ten days (Fig.~\ref{fig:final_2012_sensitivity:10days}) and one
% day (Fig.~\ref{fig:final_2012_sensitivity:1days}). It is important to note
% that this is not the sensitivity achieved by the Neutrino Triggered Target of
% Opportunity Program as the sensitivity plotted here requires precise knowledge
% of the flare time window. Compared to Fig~\ref{fig:ntoo:evtsel:neutrinosel:cutopt},
% which depicts the sensitivity for the $2011/2012$ dataset, only small changes
% in the region $\cos\theta_{\text{MPE}}<-0.8$ are visible.  
% 
% \begin{figure}[!htnp]
%   \centering
%     \includegraphics[width=0.75\textwidth]{img/final_2012_upgoing_sensitivity_10days}
% \caption[Sensitivity for 10 day flare]{Sensitivity to neutrino
% flares with a spectral index $\gamma=-2$ and duration  $10$ days as a function of $\cos
% \theta$.}
% \label{fig:final_2012_sensitivity}
% \end{figure}

\section{Data stability monitoring}
\label{data_stability_monitoring}

A dedicated monitoring system was implemented to minimize the rate of
false alerts caused by problems with the detector itself, the data
acquisition (DAQ) or the filtering software. IceCube has very
extensive  DAQ monitoring, and processing results which are available with a certain delay after data-taking.  However, the monitoring does not provide
information on the detector stability with high granularity but
declares a whole run, with a usual duration of eight hours, as either
\textit{good} or \textit{bad}. Problems such as a few strings of the
detector dropping out of the data taking shortly before the end of a
run do not render the data taken up to that point invalid. To ensure
that alerts are issued during stable running conditions, a simple but
effective online stability monitoring scheme has been developed.  The
scheme is based on the continuous monitoring (in 10 min time  bins) of
several trigger and filter event rates, representative of different event
topologies from atmospheric muons and neutrinos.
 
\subsection{Rate measurements and data quality assessment}
 \label{sec:ntoo:stab:measure}

The trigger rates of the detector, and the filter event rates of the online
filters, are quantities that are both sensitive to problems affecting
the data quality and simple to measure, record and evaluate. Trigger
event rates (e.g.~the SMT8) are sensitive to
low-level problems, such as possible errors in the trigger configuration
or an incorrect DOM calibration.  Filter event rates can also be affected by
these issues but, additionally, they give information about the
stability of the filtering chain. Problems that affect event
reconstruction or distributions of cut variables used in a filter
would also change the corresponding filter event rate.
\par All trigger and filter event rates are measured by a central
server using a dedicated software
module.  Events are counted in time bins of $600$ seconds and the
corresponding rates and time-bin meta data (e.g. start and end of the
time bin) are inserted into a relational database.
\par This database is mirrored to the Northern Hemisphere to be easily
accessible for offline studies. Storing the data in a relational
database makes it convenient to retrieve any trigger and filter event rate
for longer time scales such as hours or days. For each of the trigger and
filter event rates approximately $5 \cdot 10^4$ measurements are recorded in
the database in a full year.
\par The NToO selects $\nu_{\mu}$-induced muon tracks to detect
time-variable point sources of neutrinos. Any problem that affects the
detection and reconstruction of these muons would therefore have an
impact on this program. Thus the inputs derived from the rate
monitoring for the NToO should be related to the muon-related triggers
and filters that form the basis of the neutrino event
selection. The following trigger rates, filter event rates and ratios are
used to check the stability:

\begin{itemize}
\item {\it  Simple Multiplicity Trigger} event rate 
\item  {\it Muon Filter} event rate
\item {\it Online Level $2$ Filter} rate 
\item Ratio of {\it Online Level $2$ Filter} event rate to {\it Muon Filter} event rate
\item Ratio of {\it Online Level $2$ Filter} event rate to {\it Simple Multiplicity Trigger rate}
\end{itemize}

A combination of these rates and ratios  form a \textit{stability score}, which
will be described in Sec.~\ref{sec:ntoo:stab:stability_score}. 

As the final neutrino event selection is performed in a different
subsystem, the final-level event rate is not recorded in the
database. Due to the very low atmospheric-neutrino rate of about 2 mHz
at the final cut level, the statistical error on the rate measurement
with the default time binning of $10$ minutes would be very
large. Recording this rate with a different binning and combining it
with the other rates would make the system much more complicated.
Therefore, the final neutrino level event rate is not used as an input
in the rate-based detector stability monitoring.

\subsection{Stability-score calculation}
\label{sec:ntoo:stab:stability_score}
\begin{figure}[!tbp]
\centering
\includegraphics[width=0.6\textwidth]{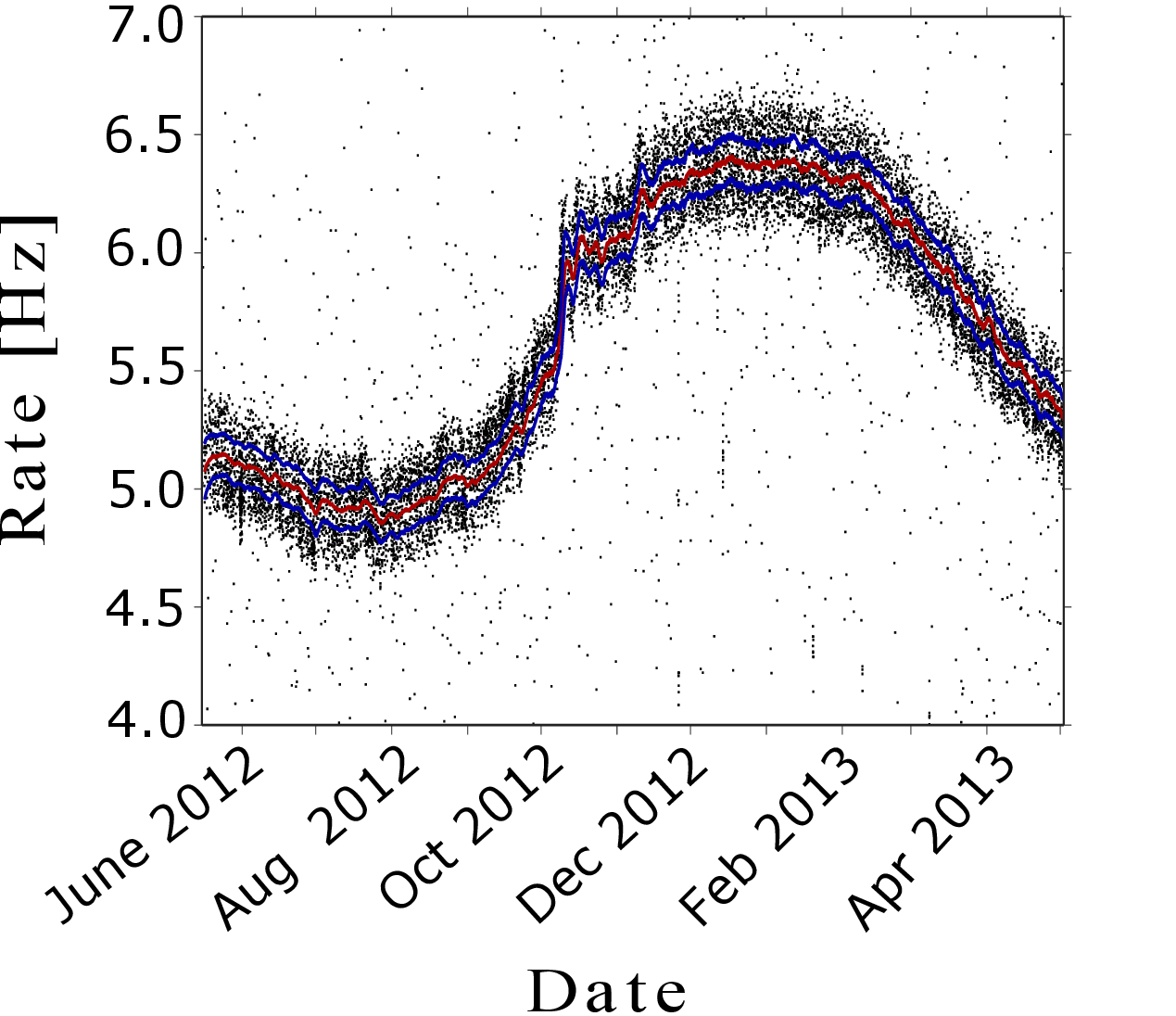}
\caption[{\it Online Level 2 Filter} event  rate with exponential moving average]{IceCube rate of events
    passing the {\it Online Level 2 Filter} over the complete IC-2012 
    season. The solid red line depicts the moving average of the {\it Online Level
    2 Filter} event rate; the blue lines show the $1\,\sigma$ exponential standard deviation around the average. Each dot corresponds to a 10 min time bin.}
\label{fig:ntoo:stab:onlineL2_with_average}
\end{figure}

The atmospheric-muon rate depends on the development of the air shower
and thus on the atmospheric density profile.  As seasonal temperature
changes of the atmosphere influence this density profile, the
atmospheric-muon rate measured in the detector shows a pattern of
seasonal variation, see
Figure~\ref{fig:ntoo:stab:onlineL2_with_average}. On top of these slow
seasonal variations, changes in the IceCube trigger rate on the
time scale of hours to days are apparent. This background of atmospheric
muons dominates all trigger and filter event rates used for the online
stability monitoring. Therefore, a simple stability decision based on
the deviation from fixed reference rates cannot be used.

A common method to predict a time series of (potentially noisy) measurements is a moving-average filter. The filter smooths noisy data to either produce smoothed data for
presentation purposes or to make forecasts of the time series. Three different averaging methods are
usually employed, simple moving averages, weighted moving averages or
exponential moving averages.
\par
An $N$-period simple moving average weights the last $N$ measurements equally to
produce the smoothed prediction. In doing this, the average always
lags sudden changes in the data. This can be overcome by applying a weight to
each measurement in the averaging process, depending upon how long ago the measurement was
taken. This requires two inputs, the number of measurements $N$ to average
over and the weight function. In the case of the stability monitoring one
would assign higher weight  to more recent measurements so that 
 recent measurements such that the average reacts faster to changes of the rates caused by a changing muon rate.

Another way to achieve this fast adaptation is an exponential moving average.
Given measurements of a quantity $x$ (e.g.~a filter event rate) at time steps $i$ (denoted as $x_i$)
the exponential moving average $S$ at time step $i$ is calculated as

\begin{align}
S_1 &= x_1\\
S_i &= \alpha x_i + (1-\alpha) S_{i-1}\;;\;\text{for $i > 1$}\,.
\label{eq:exp_avg}
\end{align}

The parameter $\alpha$ determines how fast the weight given to past
measurements decays; higher $\alpha$ gives more weight to recent
observations and reduces the impact of past measurements more
rapidly. The step width is given by the $600$-second time-bin width of
the rate monitoring.  \par Analogously to the exponential moving
average, an exponential moving standard deviation $\sigma$ can be
defined as

\begin{equation}
\sigma_i = \sqrt{\left< x^2 \right> - S_i \cdot S_i}\,.
\label{eq:exp_stddev}
\end{equation}

Here $ \left<x^2 \right>$ denotes the exponential moving average (see Eq.~\ref{eq:exp_avg}) of $x^2$.
To update the exponential moving average only the most recent calculated value of $S_i$ is needed.
This is in contrast to the simple and weighted moving averages, where the past
$N$ data points need to be kept for updates of the average. Therefore, an exponential smoothing has been chosen in the
stability monitoring  in order to  simplify the implementation of 
the moving-average calculation.

 \subsection{Implementation of the stability-score calculation}

The stability score provides a metric to compare the current detector trigger and filter event rates
in time bin $i$ to an exponential moving average of these rates up to that point in time.
The averages and standard deviations are calculated for the filter event  rates and ratios 
with the parameters $\alpha = 0.01\,$ \footnote{Until $25$ November
$2012$ $\alpha=0.005$, which gave more weight to past measurements. In order to be
better able to cope with fast rate variations due to weather changes the value of $\alpha$ was changed to $0.01$.}.
In order to judge the detector stability in a time bin $i$, a combined score $\xi_i$ is calculated as
\begin{equation}
\xi_i = \sum_j \frac{|x^j_i - S^j_{i-1}|}{\sigma^j_{i-1}}
\label{eq:badness}
\end{equation}
where $j$ enumerates the filter  event rates and ratios and $S_{i-1}$ and
$\sigma_{i-1}$ are the exponential moving averages and standard
deviation, respectively,  prior to the time bin $i$. If $\xi_i$ is below a certain
threshold $\xi_{\text{thresh}}$ the time bin $i$ is considered to be
of good quality and the averages and standard deviations are updated
according to Eq.~\ref{eq:exp_avg} and Eq.~\ref{eq:exp_stddev}. If
$\xi_i$ is above the threshold the data quality in this time bin is
judged to be bad. In this case, all final-level events in that time bin
are discarded, the time bin is counted as detector dead time and the
averages and standard deviations are not updated with the rates from
time bin $i$.

The threshold employed in the NToO is
$\xi_{\text{thresh}} = 8\,$.  
For this threshold, comparisons of the online stability monitoring with the more extensive offline quality checks  show that the
online system reliably identifies unstable detector conditions.

As an example, for IC-2012 the data taking season started 
on 15 May 2012 at 10:05:48 UTC and ended on 2 May 2013 
at 09:48:49 UTC. Of the 351.98 days between the season start and end, 322.17 days are marked as good by the stability monitoring. This results
in an uptime fraction of 91.5\%.  Typical IceCube offline analyses
for this season report an uptime fraction of around 95 \%.
\section{Technical design of the alert system}
\label{technical_GFU_design}

\begin{figure}[!htbp]                                                                                                                                                     \centering                                                                                                                                          
   \includegraphics[width=0.6\textwidth]{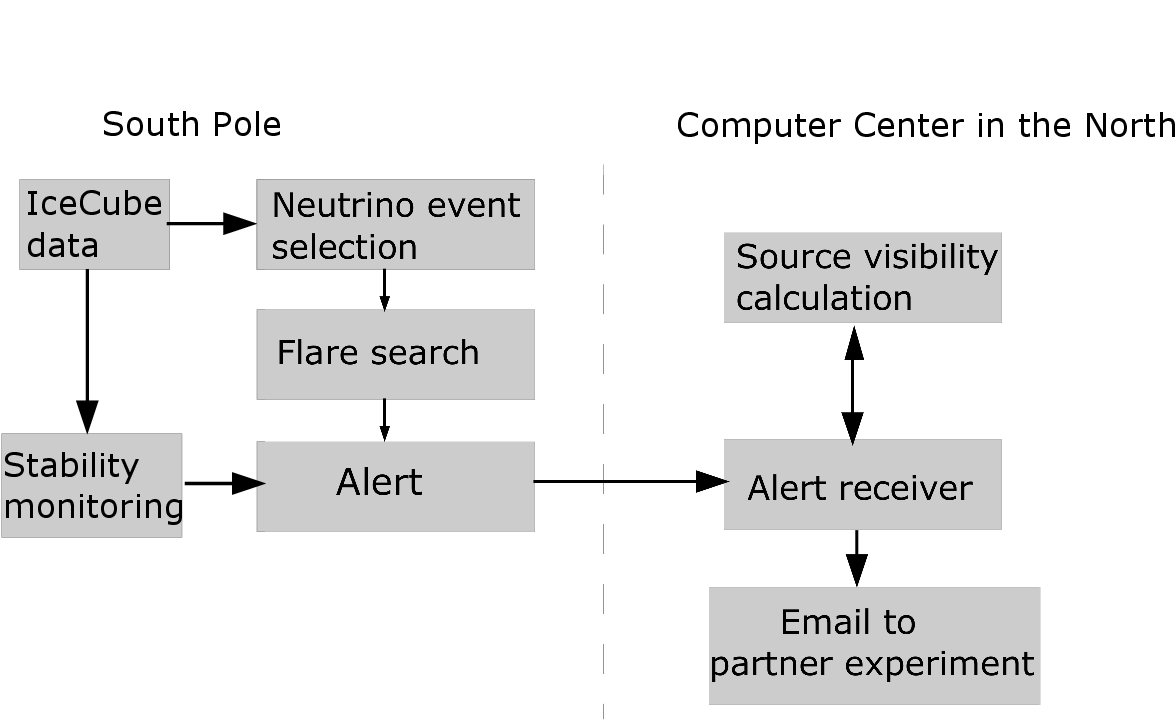}                                                                                         
 \caption{Schematic of the design of the IceCube NToO.}                                                                    
 \label{fig:ntoo:tech_design:schematic}                                                                                                                        
\end{figure}

The NToO system runs online at the South Pole with minimal human
intervention. In order to maximize the uptime of the system it has to
be very stable. The main design driver was that the failure of any of
the sub-components should not lead to the loss of the online program's data.
Therefore all components have been separated as much as possible and intermediate
results are stored frequently. The basic components of the NToO are depicted in Figure~\ref{fig:ntoo:tech_design:schematic}.

In the first step, the selection of neutrino candidate events happens inside the IceCube data-processing system
at South Pole. Each event is serialized to the text-based and human-readable
JSON (JavaScript Object Notation) format and written to a dedicated directory
on disk. The event directory is checked for new events every $30$ seconds by
the daemon that runs the time-clustering algorithm. This daemon keeps a list
of events detected in the last $21$ days from each of the monitored
sources and adds new events to the appropriate list if the detector was stable
when the event was detected. For each new event that falls into the search bin
of one of the monitored sources the time-clustering algorithm for that
particular source is run.

If the significance for an evaluated event cluster
exceeds a certain threshold (see below), an alert message containing the
source name, event positions, event times and the significance of the cluster
is generated. The alert message is then sent to the University of Wisconsin
via the IceCube Teleport System (ITS) which uses the  network of Iridium satellites. This
low-bandwidth connection  allows  short
messages  to be sent from the South Pole without any significant delay. Once the message
arrives in the North it is checked to see whether it represents a \textit{real
alert} or a \textit{test alert} from a monitoring source (see next section for an explanation of the
difference). If it is a \textit{real alert}, the
alert is forwarded to the respective partner experiment, MAGIC or
VERITAS or to both of them if the alert significance 
is above the  threshold  for MAGIC and VERITAS. Currently the alerts are forwarded via email and follow-up
observations are initiated manually. The total time delay between the (latest) neutrino event detected by
IceCube and the moment that alert is forwarded to the partner experiment is on average 12 minutes.

%{\it  {\bf Comment from MarkusV:}. Monitoring:
%
%I would start a new section with "The NToO system it is based on a
%time-clustering approach and it looks..."
%This is not about monitoring alerts anymore, but describes the running
%system and the real physics alerts.}

\section{Monitoring of alert system}
The low rate of accidental background alerts from atmospheric neutrinos (see
Figure~\ref{fig:ntoo:alert:bgalert}) makes it necessary to add additional monitoring to the
system in order to ensure that all components are working as expected. Ideally,
this monitoring should cover the whole chain, from the event selection and
stability monitoring, to the generation, sending and receiving of
alerts. In order to reach this goal, so-called test alerts can be generated at the South
Pole using the same event sample as used by the NToO.
To achieve a sufficiently large rate of test alerts
the number of source positions that are monitored should be high. Thus, 1000 random positions were chosen
as test sources, with
a flat distribution in $\cos \theta$. The threshold
for sending a test alert should be lower than the corresponding threshold
for the physics alerts in order to achieve a high number of test alerts. Thus,
the threshold for test alerts was set to $p_{obs}=0.1$ (see
Eq.~\ref{eq:ntoo:nobs}).
\par
Using the same original neutrino event sample for both the physics alerts and for the test alerts would unblind additional positions in
the sky. The usual way to test point-source analyses in a way that
preserves blindness is through scrambled data sets. The event times
are shuffled and new sky coordinates are calculated for each event.
Due to the location of the IceCube  exactly at the geographic South
Pole, only the right ascension is affected by this procedure. In the
case of the NToO, however, a continuous stream of events must be shuffled 
while preserving properties such as the azimuth and time distribution of the events.

\begin{figure*}[!htbp]
\centering
    \subfigure[tight][Spatial event distribution] {
    \label{fig:monitoring:test_alert:events}
    \includegraphics[width=0.48\textwidth]{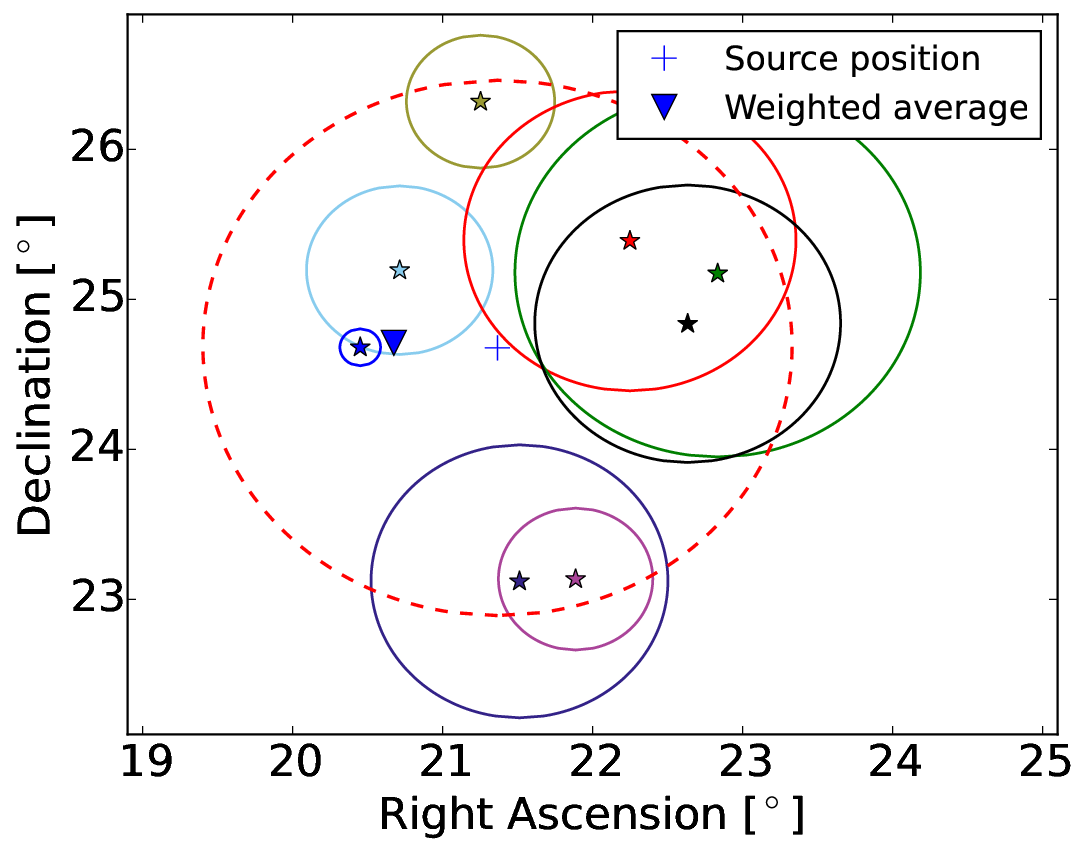}}
    \hspace{-0.5em}
    \subfigure[tight][Temporal event distribution] {
    \label{fig:monitoring:test_alert:times}
    \includegraphics[width=0.48\textwidth]{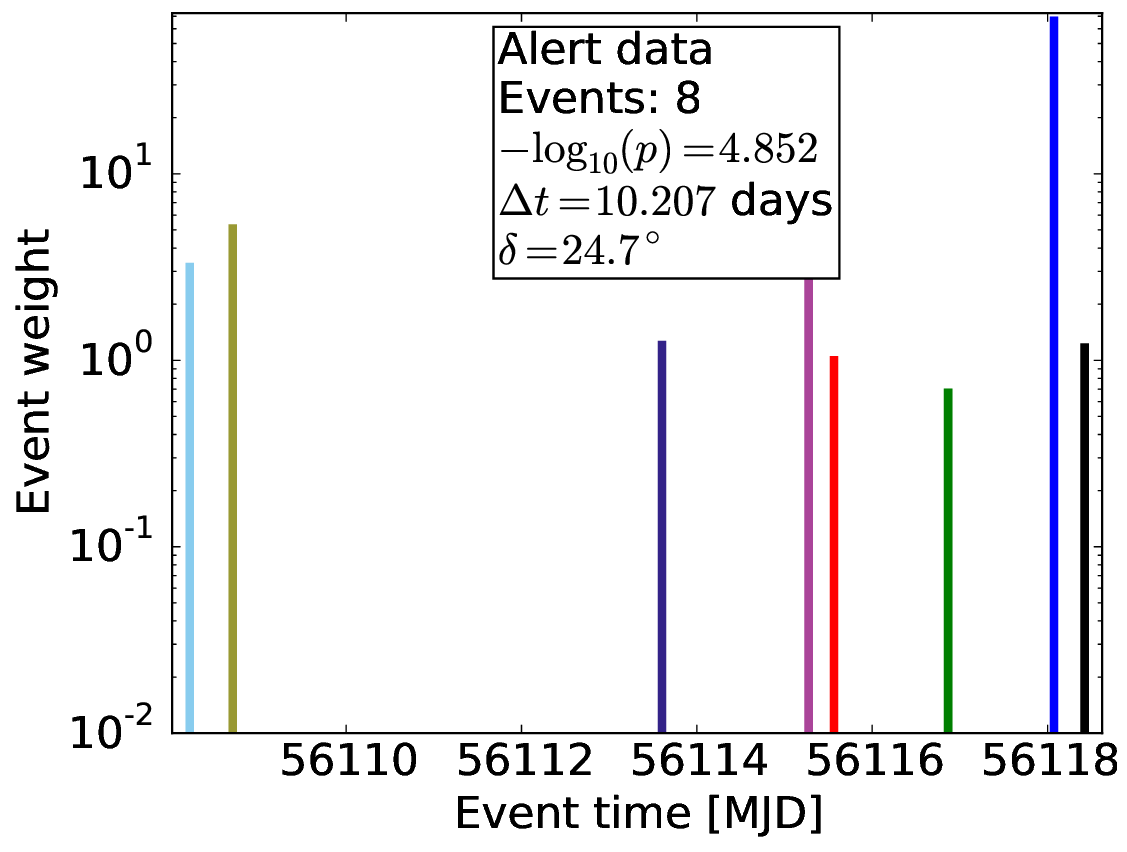}}
    \caption[Test alert]{Left panel  shows spatial  distribution of
    IceCube events (marked by stars) contributing to a test alert. The circles describe the estimated angular error
    for the reconstructed tracks. The dashed circle indicates the size of the on-source bin. Right panel shows the temporal distribution of eight events  depicted  in the left panel. The height of the bars in right panel corresponds to the event weights derived from the angular
reconstruction uncertainties. The weights are used to calculate the
weighted average direction of the events, shown as an inverse full triangle in the left panel.
 }
    \label{fig:example_alert}
\end{figure*}

% \begin{figure*}[!htbp]
%   \centering
%     \subfigure[tight][Number of test alerts per day] {
% \label{fig:monitoring:nalert_hist}
%     \includegraphics[width=0.5 \textwidth]{img/alert_rate_hist.eps}}
%     \hspace{-1.5em}
%     \subfigure[tight][Time between test alerts] {
% \label{fig:monitoring:deltat_hist}
%     \includegraphics[width=0.5 \textwidth]{img/deltat_hist.eps}}
% \vspace{-0.4cm}
%     \subfigure[tight][Number of events in alert] {
% \label{fig:monitoring:nevent_hist}
%     \includegraphics[width=0.5 \textwidth]{img/nevent_hist.eps}}
%     \hspace{-1.5em}
%     \subfigure[tight][Significance of test alerts] {
% \label{fig:monitoring:significance_hist}
%     \includegraphics[width=0.5 \textwidth]{img/significance_hist.eps}}
%     \caption[Alert system monitoring plots]{Monitoring information derived
%     from the test alerts for the NToO. See text for a description.}
% \label{fig:monitoring}
% \end{figure*}
 
To randomize the event coordinates in right ascension for the
blind generation of test alerts one could, in principle, assign each event a random azimuth
angle. This would, however, destroy the pattern due to the azimuth-dependent efficiency
of the detector, see~Figure~\ref{fig:neutrinosel:azi_asymm}. In order to preserve this pattern in the scrambled dataset, the
conversion of local coordinates (zenith and azimuth) to sky coordinates (right ascension
and declination) for each event is done not with its original event time, but with the time of the
previous neutrino event. The first event after the startup of the event-selection process is
assigned a random right ascension. As the rate of atmospheric-neutrinos is about 2 mHz this
results in a random shift of each event by several degrees on average.
%  \begin{figure*}[!htbp]
% \centering
%     \subfigure[tight][Spatial event distribution] {
%     \label{fig:monitoring:test_alert:events}
%     \includegraphics[width=0.40\textwidth,height=6cm]{img/testalert_2012_events.eps}}
%     \hspace{-1.5em}
%     \subfigure[tight][Temporal event distribution] {
%     \label{fig:monitoring:test_alert:times}
%     \includegraphics[width=0.40\textwidth,height=6cm]{img/testalert_2012_times.eps}}
%     \caption[Test alert]{Spatial (Figure~\subref{fig:monitoring:test_alert:events}) and
%     temporal (Figure~\subref{fig:monitoring:test_alert:times}) distribution of
%     events contributing to a test alert. The circles describe the estimated angular error
%     for the reconstructed tracks. The dashed circle indicates the size on-source bin.
% The height of the bars in Fig (b) corresponds to the event weights derived from the angular
% reconstruction uncertainties. The weights are used to calculate the
% weighted average direction of the events, the inverse full triangle in Fig (a).
%  }
%     \label{fig:example_alert}
% \end{figure*}
% \subsubsection{Monitoring Web Page}
% \label{sec:ntoo:alert:monitoring:webpage}
\begin{figure*}[!htbp]
  \centering
    \subfigure[tight][Number of test alerts per day] {
\label{fig:monitoring:nalert_hist}
    \includegraphics[width=0.45 \textwidth]{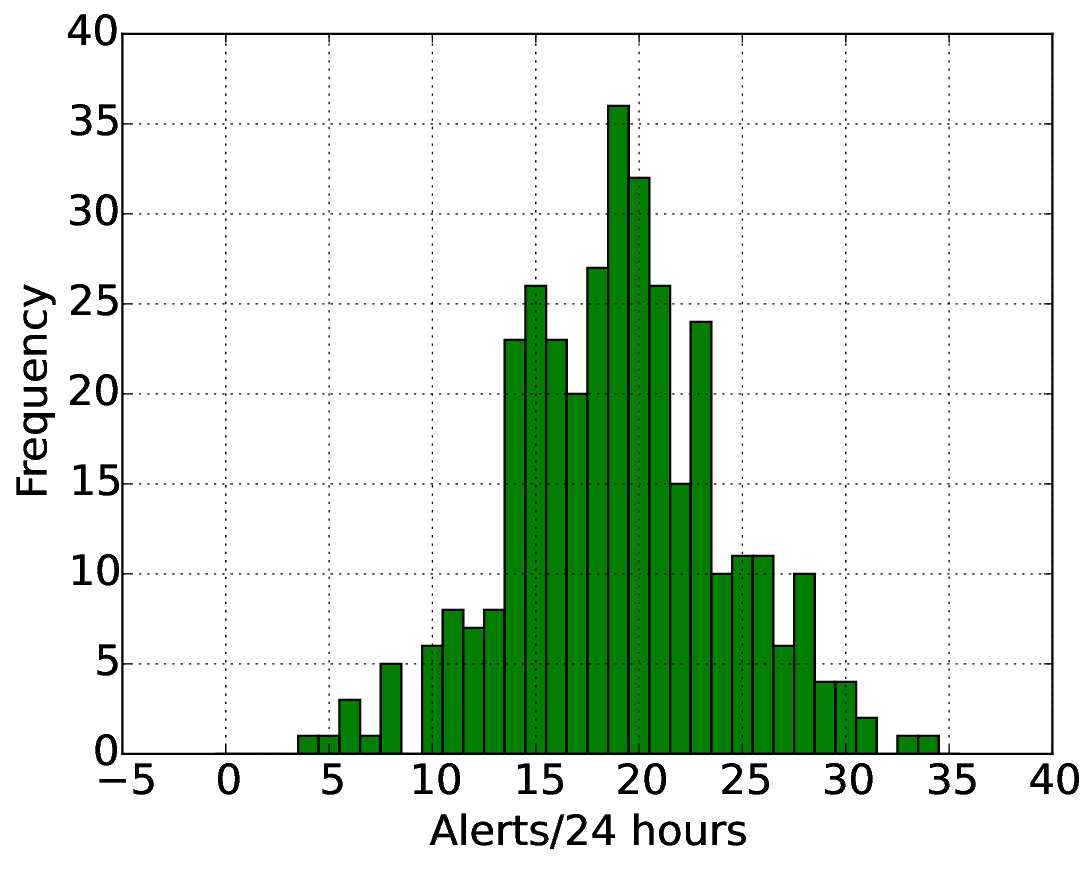}}
    \hspace{-0.5em}
    \subfigure[tight][Time between test alerts] {
\label{fig:monitoring:deltat_hist}
    \includegraphics[width=0.45 \textwidth]{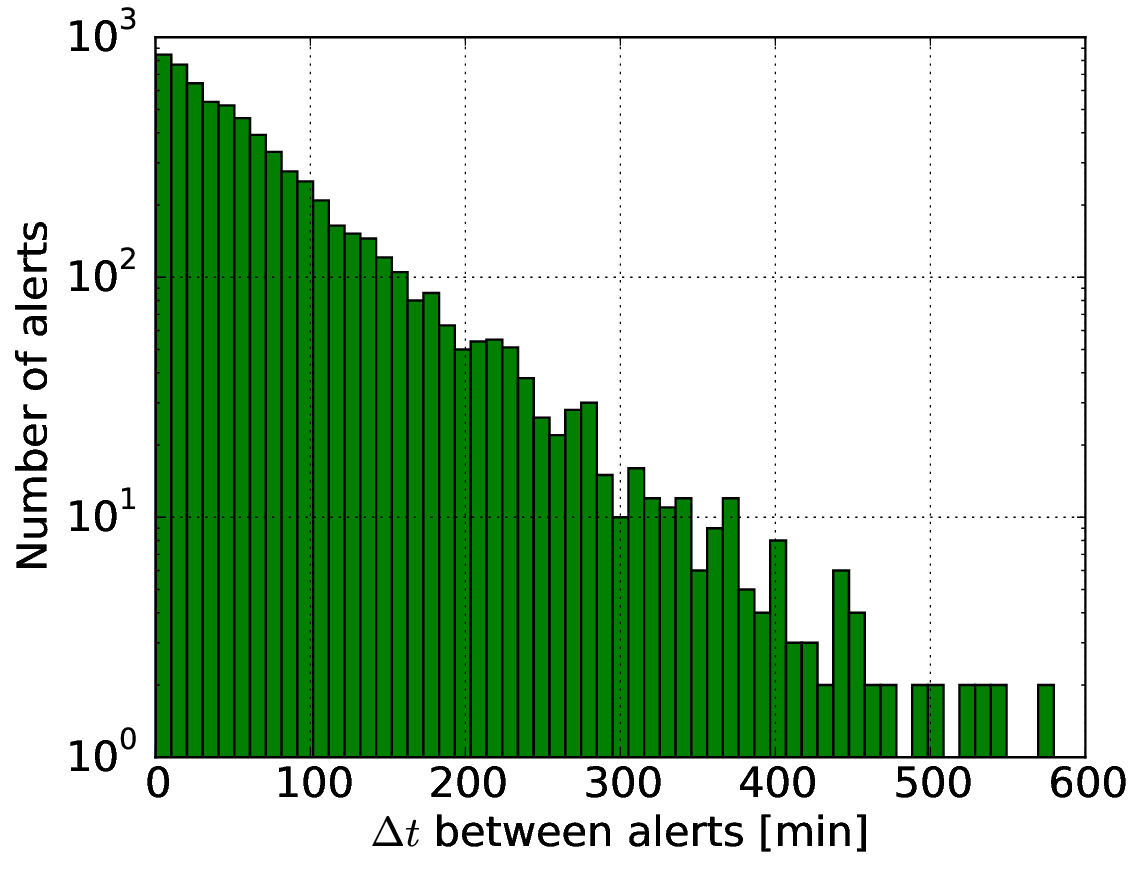}}
\vspace{-0.4cm}
    \subfigure[tight][Number of events in alert] {
\label{fig:monitoring:nevent_hist}
    \includegraphics[width=0.45 \textwidth]{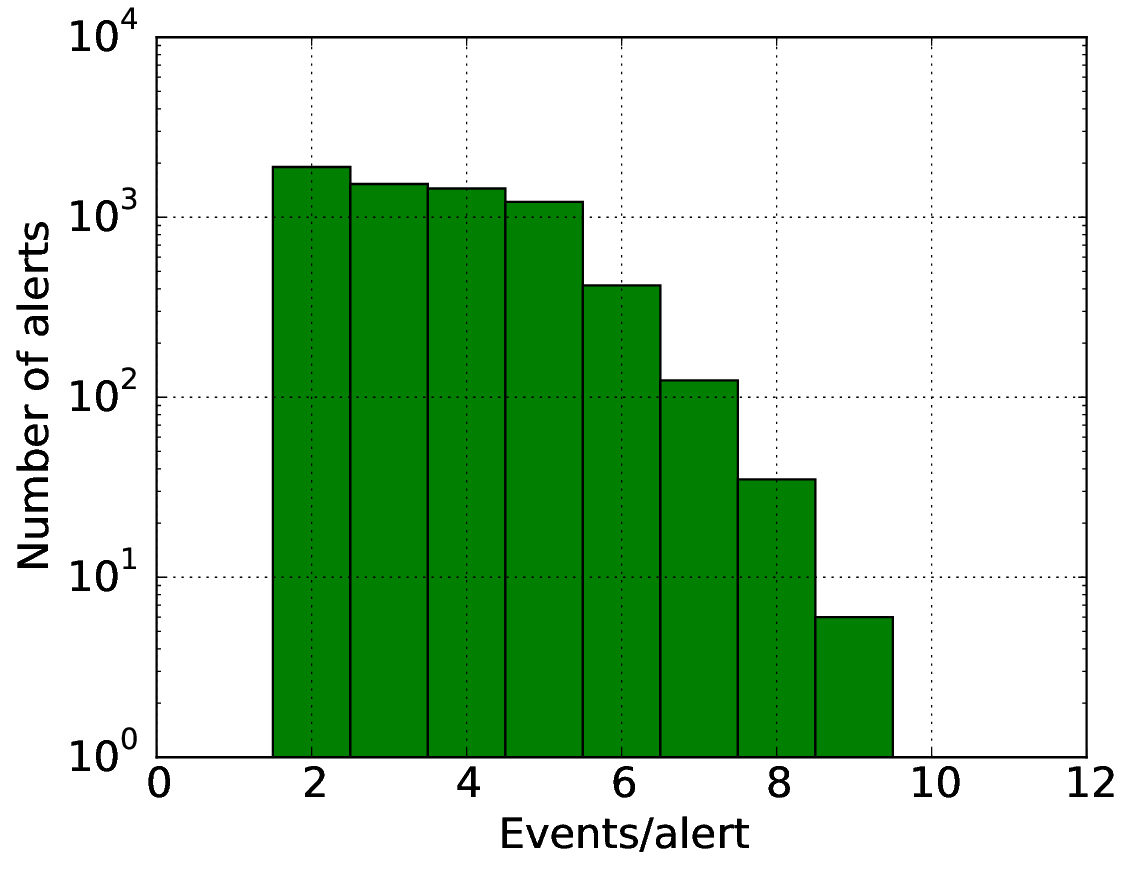}}
    \hspace{-0.5em}
    \subfigure[tight][Significance of test alerts] {
\label{fig:monitoring:significance_hist}
    \includegraphics[width=0.45 \textwidth]{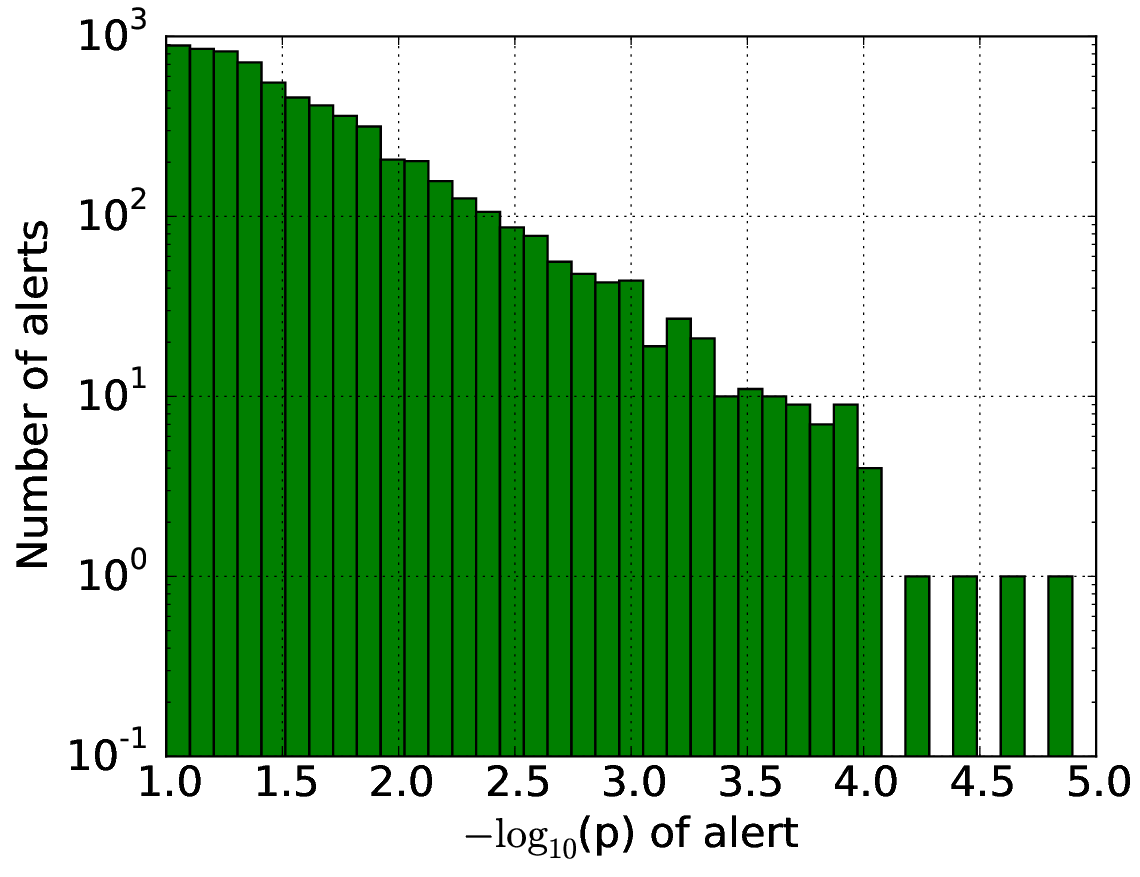}}
    \caption[Alert system monitoring plots]{Monitoring information derived
    from the test alerts for the IceCube NToO. See text for a description.}
\label{fig:monitoring}
\end{figure*}

The test alerts generated from the blinded event sample are collected
and analyzed. To aid the interpretation of these alerts a web page was
created that displays each alert. The web page is automatically
updated upon receiving a new test alert. In addition to each
individual test alert, the global properties of all test alerts
received to date are shown, e.g.~the rate of test alerts, their zenith
distribution and their significance distribution. Each alert is
displayed on a web page showing the distribution of events
contributing to the alert both in time and space (see
Figure~\ref{fig:example_alert}). In the case of an alert for an
astrophysical source this allows for a rapid inspection of the event
properties.

As an example,  Figure~\ref{fig:example_alert} depicts a high-multiplicity test alert,
consisting of $8$ events, issued on $7$ July $2012$. It corresponds to the test alert with
the highest significance in the IC-2012 data taking season with
$-\log_{10}(p_{obs})=4.85$. The contributing events were detected
over a duration of $10.2\,$days.
\par
For each alert the weighted-average direction is calculated as
\begin{equation}
x_{avg} = \sum_i \frac{\sum_j \sigma_j^2}{\sigma_i^2} x_i
\label{eq:evtweight}
\end{equation}
where the $\sigma_i$ are the resolution estimates of the individual events (i.e. {\it Cram\'er-Rao Resolution Estimate}) and the $x_i$ are their directions described by the right ascension and declination.
The weighted average is displayed as a full triangle  in the spatial event plot, the individual
event weights ($1/ (\sigma_i^2/ \sum_j \sigma_j^2)$) are represented as the height of the bars in the temporal plot.

Plots of the global properties of all monitoring alerts received to date can be used
to monitor the stability of the operation of the whole alert system. For example, changes in
the total test-alert rate can indicate problems with the event selection or uptime calculation. Long delays between the detection of the events and the arrival of the test alerts in the North
can be a sign of problems with the data processing, the stability-monitoring database or 
the transfer of  test alerts to the North. Figure~\ref{fig:monitoring} shows some of the quantities
derived from the test alerts which allow the alert system to be monitored.

An important quantity to monitor is the rate of received test alerts
(Figure~\ref{fig:monitoring:nalert_hist}). The regular arrival of test alerts in
the North is used as a ``heartbeat'' for the overall system. If no test alert is
received for more than six hours, a warning email is
issued to a list of people  so that the cause can be investigated. Warning emails are reissued
every two hours if no new alert has been received in the meantime. This
threshold of six hours for warning emails is rather conservative, as can be
seen in Figure~\ref{fig:monitoring:deltat_hist}. This figure shows the histogram of the wait
times between subsequent test alerts. It follows the expected exponential
distribution reasonably well. A  time difference  of six hours is well within the range
of expected waiting times. However, to enable timely interventions, an early
warning is preferred. Figure~\ref{fig:monitoring:significance_hist}
depicts the distribution of the significances of the test alerts.

% \begin{figure}[!t]
%  \centering
%  \includegraphics[width=3.0in]{threshold.eps}
%  \caption{\small Expected number of accidental background alerts per year for a source
%  at different  declinations as a function of the alert threshold expressed
%  in units of standard deviations corresponding to a one-sided p-value.}
%  \label{alertrates}
%  \end{figure}
 
\section{Results of  NToO program}
\label{results}

{ The IceCube follow-up programs such as  optical, X-ray follow-up and NToO have been running in a stable fashion
for a few years and are taking high-quality data from both IceCube and the follow-up instruments. 
The results are  the subject of a forthcoming publications. Only a short status report will be given here, highlighting the most important results.

For the optical and X-ray follow-up, no significant excess of multiplets was found  since the inauguration of the program in December 2008. One neutrino triplet was found in the data in February 2016, this result will be subject of a forthcoming IceCube publication. In March 2012, the most significant  alert during the first three years of operation of the optical and X-ray follow-up program was issued by IceCube. In the follow-up observations performed by the PTF, a Type IIn supernova PTF12csy was found
$0.2^{\circ}$ away from the neutrino alert
direction~\cite{snptf12csy}. The supernova has a redshift of $z=0.0684$,
corresponding to a luminosity distance of about 300 Mpc, and the
Pan-STARRS1 survey shows that its explosion time was at least 158 days
(in the host-galaxy rest frame) before the neutrino alert, implying that a
causal connection is unlikely~\cite{snptf12csy}.

%The NToO system is based on a time-clustering approach and looks for any time frame  with a significant deviation 
%of the number of detected neutrinos from the expected background.
\begin{table*}[h!bt]
\centering
\small
\begin{tabular}{|c||c|c|c|c|c|c|c|c}\hline
alert  & Source&Time  & $-\log_{10}(p_{\text{obs}})$  & $N_\text{{obs}}$ & Duration  & Follow-up & Observed \\
ID & &(UTC) &  &  & (days) &Instrum.& yes/no \\
\hline
1 & PG 1424+240$^{*,**}$ & 2012-04-14  23:47 & 3.47 & 6& 7.617 & No &- \\
2 &GB6 B1310+4844 &2012-08-20 09:53 & 3.75 & 6& 6.344 & No &- \\
3 & 4C15.54&2012-09-13 01:52 & 4.06 & 2& 0.001 & MAGIC& No \\
4 &SBS 1150+497 $^{**}$&2012-11-09 07:28 & 4.64 & 6& 4.169 & VERITAS& Yes \\
5 &RGB J0152+017$^{*}$ &2013-04-29 06:36:& 4.07 & 8& 15.801 & No & -\\
6 & RGB J0505+612$^{**}$&2013-09-12 20:00 &3.31 (4.10)& 7 (10)& 11.790 (20.73)&  MAGIC& Yes\\
7 &1ES 2344+514$^{*}$ &2014-02-19 23:18 & 4.07 (4.23) & 8 (9)& 12.844 (16.40)&  VERITAS& Yes\\
8 & 1ES 1959+650$^{*}$&2014-03-09 10:28 & 3.40 & 9& 20.944 & MAGIC& No \\
9 &B3 1708+433 &2014-06-22 02:42 & 4.34 & 3& 0.118 & No & -\\
10 &PKS 1717+177$^{**}$& 2014-09-24 13:47 & 3.20 & 2& 0.007 & No& - \\
11 & MG4~J200112+4352$^{*}$&2014-10-05 15:05 & 4.05 & 9& 18.631 & VERITAS& Yes \\
12 &B3 1343+451 &2014-11-16 17:00 & 3.64 (5.04) & 3 (4)& 0.301 (0.576) &  VERITAS&No \\
13 &AO 0235+164 $^{**}$&2015-04-27 04:55 & 3.97 & 8& 16.395 & No &-\\
14 &CGRaBS J0211+1051 $^{**}$&2015-07-05 00:06 & 4.09  &4  &  1.205  & VERITAS&No\\
\hline
\end{tabular}
\caption[Alert data for 2012/2015 alerts]{Overview of the IceCube alerts generated by the NToO up to 31 December 2015;  see text for more details.Time of alert corresponds to the time when alert was received at North. (Follow-up Instrum.= Follow-up Instruments;  Numbers in brackets correspond to the followed alert during
 the next few days, $^{*}$ - known VHE source, $^{**}$ - existing VHE limit in \cite{veritasblazar}).	}
\label{tab:ntoo:alerts}
\end{table*}

From the inauguration of the NToO program, on 14 March  2012, to 31 December 2015, 14 alerts were sent: 4 in 2012, 2 in 2013, 6 in 2014 and 2 in 2015. The program continues, and alerts during 2016 and beyond will be reported elsewhere. From the above-mentioned 14 alerts issued, 8 of those
were forwarded and  4 (out of  8) were followed-up by MAGIC  or VERITAS observations.  
Another  6  alerts (out of 14 issued) were not forwarded due to bad observing conditions  or the partner experiment was not operational. 

Table~\ref{tab:ntoo:alerts} gives an overview of all of the alerts
generated by the NToO up to  31 December  2015. Below, only the alerts forwarded to
the partner experiments  are  discussed in more detail.

The most interesting alert (alert \#4 in Table~\ref{tab:ntoo:alerts} )
was generated on $9$ November $2012$, consistent with position  of the source
SBS~1150+497 (located at zenith angle $\theta=139.5^\circ$, with
respect to IceCube).  The alert comprised six events observed during
$4.169\,$days. The spatial and temporal  distribution of these events is shown in
Figure~\ref{fig_monitoring}.  The Poisson probability (pre-trial) for
this observation is $-\log_{10}(p_{\text{obs}}) = 4.64$,  the post-trial  probability $-\log_{10}(p_{\text{obs}}) = 2.60$, making it the
most significant alert sent during this IceCube season (IC-2012).  The alert
was forwarded to the VERITAS collaboration and resulted in a follow-up
observation. Due to poor weather and bright moonlight conditions,
VERITAS observations were not possible until 12 November 2012, at
which point the source was visible at low elevation at the very end
of the night. A further observation was made on the following night
giving a total exposure time of 71.5 min. No evidence for gamma-ray
emission was seen from the position of the source, giving an integral
flux upper  limit (99\% confidence) above 300 GeV  of $3.0 \times
10^{-10}$ cm$^{-2}$s$^{-1}$ for an assumed
differential spectrum with spectral index $\gamma=2.5$.

Another high-significance alert was sent to VERITAS on 19 February
2014, followed by a second alert on 23 February 2014, spatially coincident with
the source 1ES 2344+514 (alert \#7). The first of these was triggered
by 8 neutrinos observed over a period of 12.844 days, with
$-\log_{10}(p_{\text{obs}})= 4.07$. An additional event, observed 3.6
days later, increased the $p$-value to $-\log_{10} (p_{\text{obs}}) =
4.23$, the post-trial  probability $-\log_{10}(p_{\text{obs}}) = 2.31$,  which resulted in the second forwarded alert for this source.
The source was barely visible to VERITAS (zenith angle $> 60^{\circ}$),
and weather conditions were poor. The online VERITAS analysis showed
no evidence for gamma-ray emission (no excess was detected), 
indicating that the source flux was likely not exceptionally high above a few TeV.

During the next IceCube season (IC-2014) another two alerts were
sent to VERITAS. The first, generated on 5 October 2014 corresponded
to the source MG4~J200112+4352 (alert \#11), which had been recently reported as VHE emitter by the MAGIC collaboration~\cite{alert11}. A one-hour observation
was performed, but under extremely poor weather and bright moonlight
conditions. No conclusion regarding the gamma-ray flux state is
possible with these data. The second alert was issued by the NToO system for
the source B3 1343+451 on 16 November 2014 (alert \#12), but the
source was again barely visible (zenith angle larger than 60$^{\circ}$) and
so follow-up observations were not performed. The last alert was sent
to VERITAS on 5 July 2015 for the source CGRaBS J0211+1051 (alert
\#14), but at this time VERITAS was undergoing its annual
summer shutdown, and so no observation was made.

For MAGIC, the first alert was sent on 14 April 2012, spatially coincident with 
the source 4C15.54 (alert \#3). However, as the MAGIC telescope was in
a commissioning phase, the alert could not be followed up.  Then, a
series of four alerts were issued by the NToO from 12 September 2013 to
21 September 2013 for the source RGB J0505+612 (alert \#6). The alert
resulted in a follow-up observation by MAGIC (1 hour), which showed no
statistically significant evidence for gamma-ray emission.  The
computed integral flux  upper limit (99\% confidence) at energies
$>$200 GeV is 1.57 x 10$^{-11}$ cm$^{-2}$s$^{-1}$. The last alert
forwarded to MAGIC was generated on 9 March 2014 for the source 1ES
1959+650 (alert \#8), but the low elevation of the source precluded
observations.

\begin{figure}[!t]
\centering
\includegraphics[width=0.48\textwidth]{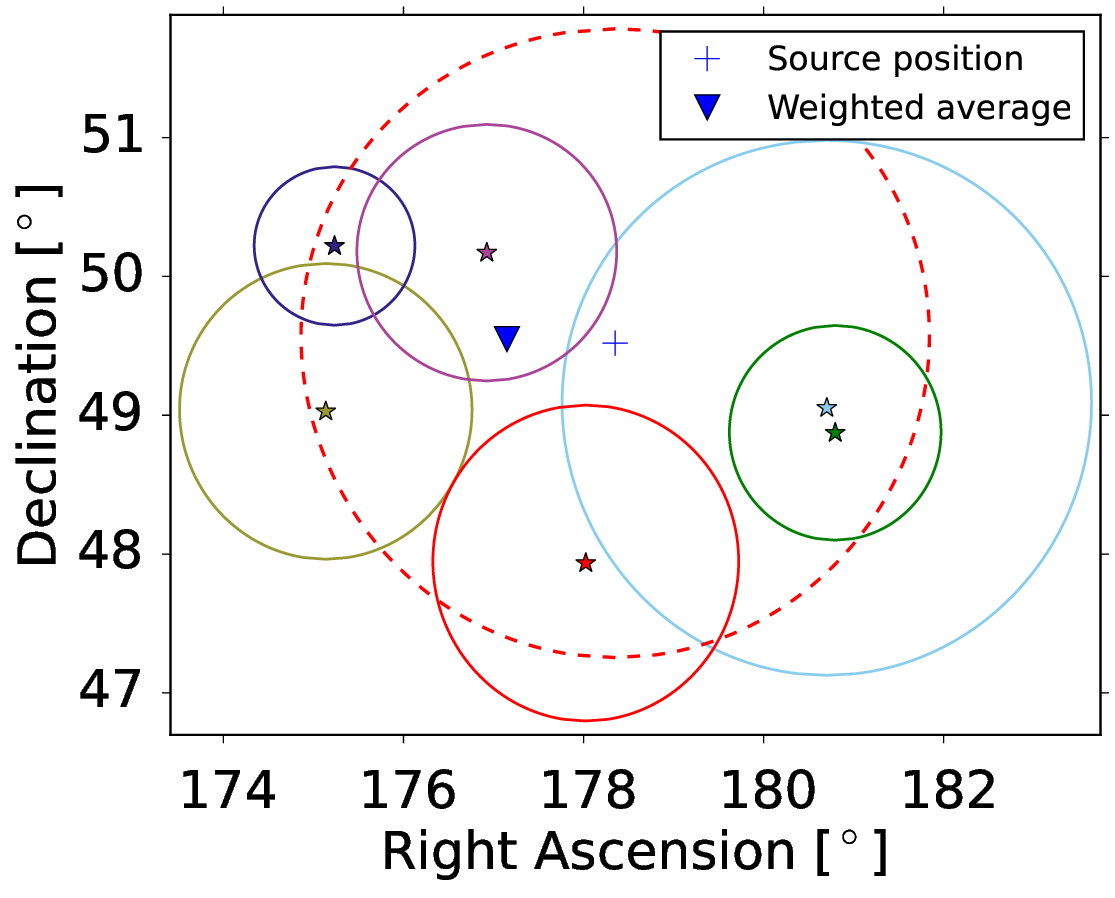}
\includegraphics[width=0.48\textwidth]{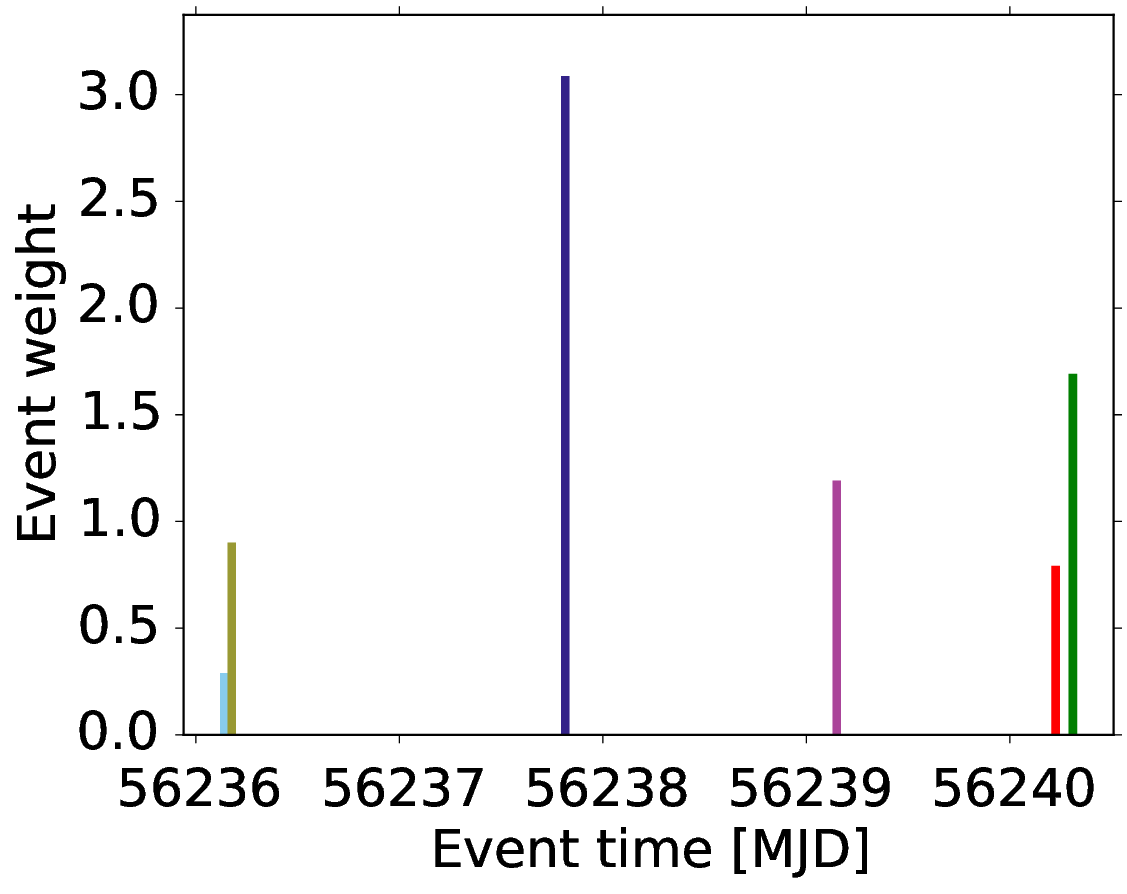}
\caption{Left panel shows position of events (star symbols) and related uncertainty (circles) from the alerts that were sent to VERITAS on 2012 November 9. The weighted average of the
contributing events is calculated using an event-by-event angular resolution
estimator. The dashed circle indicates the size of the on-source bin. Right panel shows the temporal distribution of eight events  depicted  in the left panel. }
\label{fig_monitoring}
\end{figure}

\section{Recent and upcoming improvements }

The currently deployed neutrino event selection in the NToO employs
simple cuts on a number of variables that discriminate between signal
neutrinos and the atmospheric-muon background. The cuts on these
parameters have been optimized to achieve best sensitivity.  However,
further improvements in signal and background separation should be
possible through the use of more sophisticated discrimination
algorithms, such as boosted decision tree (BDT) \cite{BDT} and
multivariate learning machines. The aim is to replace the present
NToO selection by developing a new event selection, which could also be
used by other IceCube follow-up programs. This new event
selection is comparable to offline point-source samples and will cover
the entire sky. Here, a short
description of the new BDT selection is presented, which has been
implemented for the IC-2015 data-acquisition season.

\begin{figure*}[!tp]
%\vspace{-0.8cm}
  \centering
    \includegraphics[width=0.70\textwidth]{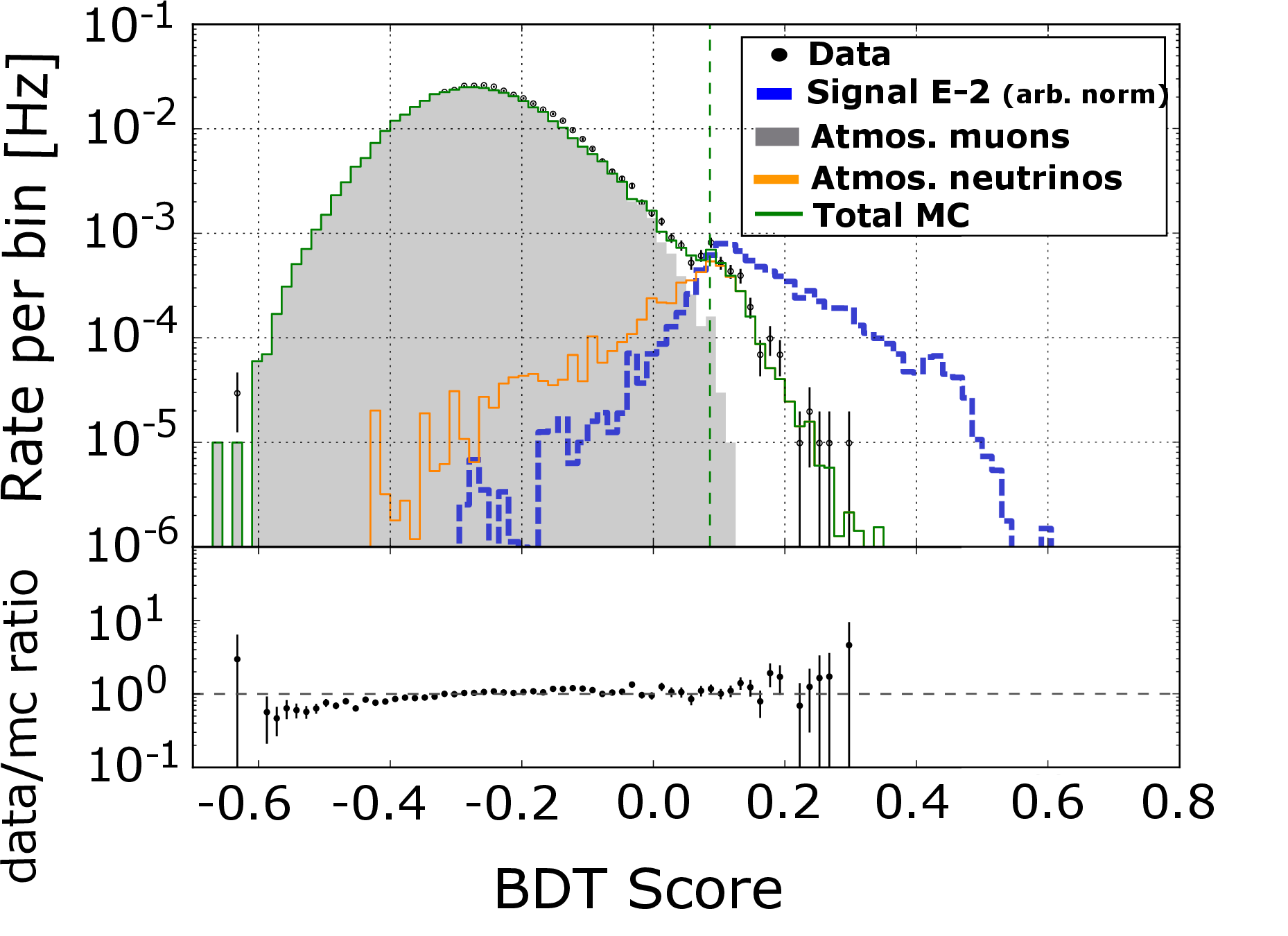}  
    
     \includegraphics[width=0.66\textwidth]{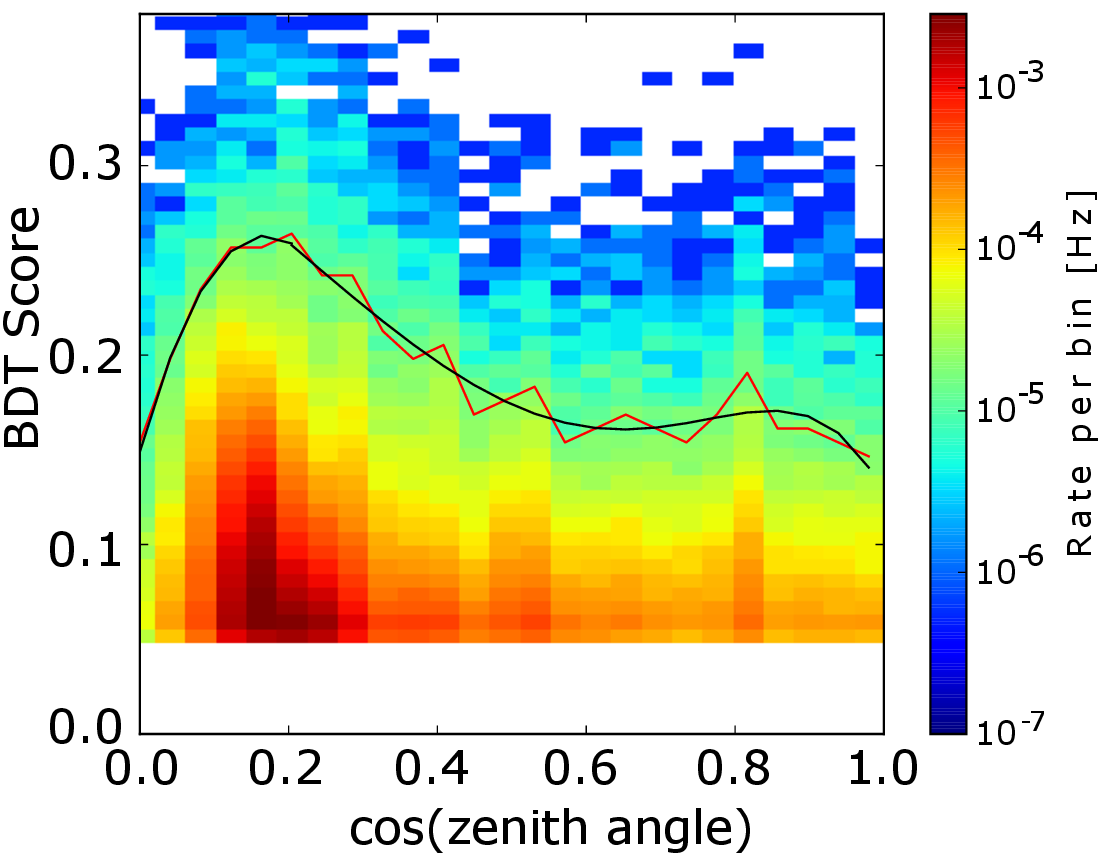}
\vspace{-0.2cm}
\caption[Sensitivity for different cut optimizations]{(Upper  panel) IceCube distribution of  BDT score  for the ensemble of trees trained with an $E^{2}$ spectrum. Vertical dashed line corresponds to the optimized BDT cut (Northern Hemisphere). (Lower panel) The BDT score as a function of cos(zenith angle). The red line keeps approximately an equal rate per zenith bin ($\simeq 10^ {-4}$  Hz). A piecewise polynomial function is then fitted to the red curve (black curve).}
\label{fig:ntoo:evtsel:neutrinosel:cutoptbdt}
%\vspace{-0.3cm}
\end{figure*}

For the new BDT selection scheme, the multivariate cuts were based on
14 observables obtained by choosing parameters with low correlation in
the background event sample, but with high discriminating power
between signal and background.  Observables specifying the geometry and
time evolution of the hit pattern, as well as the quality and
consistency of the various track reconstructions and the number of
strings with signals are used.  The BDT training was done with
simulated signal events for a soft neutrino spectrum of $E^{-2.7}$ and
for an $E^{-2}$ spectrum. As an example,
Figure~\ref{fig:ntoo:evtsel:neutrinosel:cutoptbdt} depicts results of
the BDT training for an $E^{-2}$ spectrum.  
A set of real data provided the background sample for training. Additionally,
for the simulated signal, the reconstructed track was required to be
within of $5^{\circ}$ of the simulated direction in order to train the
BDTs with only well-reconstructed events.  The final selection cut on
the BDT output variable was optimized to provide the best discovery
potential for an $E^{-2}$ neutrino flux, which results in a BDT score
value of 0.106. This final cut leads to a rate of 2 mHz for the final
sample and, as shown in Table~\ref{tab:rates:efficiency2}, to a
better signal efficiency (with respect to  {\it Online Level 2 Filter}
efficiency) than the original NToO cuts.  The BDT-based event
selection leads to an improvement in the signal efficiency of about
$+12(+18)\%$ for an $E^{-2}$ ($E^{-2.7}$) spectrum with respect to the
simple cuts.

The BDT selection was also used for the Southern Hemisphere (zenith
angle $\theta<90^{\circ}$). However, instead of a single BDT-score cut
value, a zenith-dependent cut was applied in order to select a
constant number of events per solid angle, as shown in
Figure~\ref{fig:ntoo:evtsel:neutrinosel:cutoptbdt} (Lower panel). This
zenith-dependent cut was also optimized with respect to sensitivity and
discovery potential for an $E^{-2}$ neutrino spectrum. The optimized
cut described by a polynomial fit
(Figure~\ref{fig:ntoo:evtsel:neutrinosel:cutoptbdt} (Lower panel)) leads to a
total data rate of 2.1 mHz for the South only and an average signal
efficiency of about 45 \% (with respect to the {\it Online Level 2 Filter}), see also Table~\ref{tab:rates:efficiency2}.
\begin{table*}[!t]
%\vspace{-0.5cm}
\small
\begin{center}
\hspace{-0.0cm}
\begin{tabular}{ c|c|c|c|c|c } 
{Cut Level} & Data  rate& Atm. $\nu_{\mu}$ rate & E$^{-1}$ Eff. & E$^{-2}$ Eff. & E$^{-2.7}$ Eff. \\
& (mHz)  & (mHz) & (\%) & (\%) & (\%)\\
\hline
Northern Hemisphere&   & & & &\\
Simple Cuts &  $ 2.0 $  & $1.9 $  & 79&69& 54\\
BDT $E^{-2.0}$     &  $1.9  $  & $1.7 $ & 86&81& 72\\ 
Southern Hemisphere&   & & & &\\
BDT $E^{-2.0}$     &  $2.1  $  & $0.06 $ & 78&45& 35\\ 
%BDT $E^{-2.7}$     &  $ 2.7 $  & $2.5 $ & 84&78&65 \\ 
%BDT CSW     &  $ 3.1 $  & $2.8 $  & 83&80& 72\\
\end{tabular}
\vspace{0.5 cm}
\caption{Data and atmospheric neutrino muon for different cut progression in IceCube. The signal efficiency for an $E^{-2}$ neutrino spectrum and for well reconstructed events  with $\Delta\Psi<5^{\circ}$ with respect to the {\it Online Level 2  Filter} is also shown.   }
\label{tab:rates:efficiency2}
\end{center}
\vspace{-0.5 cm}
\end{table*}
	
The first step in establishing the NToO program was to demonstrate its
technical feasibility and to prove that a time-dependent point-source
search can be run stably and reliably over long periods of time at the
South Pole. Therefore a simple search technique like the binned method
was implemented first. However, current offline IceCube searches for
neutrino point sources usually employ unbinned maximum-likelihood
methods~\cite{braun} to increase the discovery potential. Such an
approach has now also been implemented for the NToO \footnote{ At the
  moment the unbinned maximum likelihood is implemented as a
  standalone cpython module, which will be be included in the next
  forthcoming upgrade of the NToO system.}, which allows the alert  significance to be calculated by taking into account an event-by-event angular
reconstruction uncertainty estimation and an energy estimation of the
event.

Upgrading the NToO with a BDT-based event selection and a subsequent
likelihood analysis leads to an increased sensitivity in the Northern
Hemisphere of 30-40\%, yielding a comparable sensitivity to the
standard offline point-source analysis ~\cite{stasik}.  It also opens up
the possibility of observing neutrino flares in the Southern Hemisphere
and to forward these alerts to the H.E.S.S. collaboration~\cite{hess},
with whom a memorandum of understanding has been established.
 
In previous years of operation of the NToO systems, neutrino candidate
event selections, multiplet selection and alert generation all took
place within the data-acquisition system at  the South Pole.  This system
was found to be somewhat inflexible and difficult to expand. To
address these shortcomings, the NToO systems are currently
transitioning to a new approach.  Instead of selecting the neutrino
candidates at the South Pole, a BDT-selected stream of single
high-quality neutrino events is transmitted to the North via a rapid
satellite communication channel. Follow-up processes in the North now
evaluate the neutrino candidates, and generate alerts for external
observatories; see~\cite{stasik} for a more detailed description.
 Until December 2015, alerts were sent to the partner experiments privately. However,  in the future we plan to distribute alerts 
to the full multi-messenger astrophysics community (ANTARES, KM3NeT, Auger, H.E.S.S./CTA, LIGO/VIRGO, etc.)  via the Astrophysical Multimessenger Observatory Network (AMON)~\cite{amon}.

\section{Summary and Conclusions}
\label{sec:Conclusions}

In this work we described a NToO 
program, which uses  IceCube to monitor a list of predefined source candidates for neutrino flares.  
An important goal of this program was to establish and to test procedures to trigger promptly the gamma-ray community to collect  sensitive VHE data from specific sources during periods of time when IceCube measures a potential increase in their neutrino flux.
These periods of elevated emission (``flares''), both in gamma rays and neutrinos,
are of particular interest to identify the sources of astrophysical
neutrinos, and to understand the source emission mechanisms.  The
second goal of the NToO is to increase the discovery potential for
time-variable point-sources of neutrinos with IceCube. The detection
of a high-energy gamma-ray flare with an IACT triggered by an alert
from IceCube can help to establish the neutrino signal, even if it is
not significant enough on its own to qualify as a discovery. The NToO
is the first online analysis in IceCube searching for neutrino flares
from point sources on time scales longer than a few minutes.
We have shown that such an analysis can be done efficiently and reliably, and presented the first results from this program.

\acknowledgments
We acknowledge the support from the following agencies: U.S. National Science Foundation-Office of Polar Programs, U.S. National Science Foundation-Physics Division, University of Wisconsin Alumni Research Foundation, the Grid Laboratory Of Wisconsin (GLOW) grid infrastructure at the University of Wisconsin - Madison, the Open Science Grid (OSG) grid infrastructure; U.S. Department of Energy, and National Energy Research Scientific Computing Center, the Louisiana Optical Network Initiative (LONI) grid computing resources; Natural Sciences and Engineering Research Council of Canada, WestGrid and Compute/Calcul Canada; Swedish Research Council, Swedish Polar Research Secretariat, Swedish National Infrastructure for Computing (SNIC), and Knut and Alice Wallenberg Foundation, Sweden; German Ministry for Education and Research (BMBF), Deutsche Forschungsgemeinschaft (DFG), Helmholtz Alliance for Astroparticle Physics (HAP), Research Department of Plasmas with Complex Interactions (Bochum), Germany; Fund for Scientific Research (FNRS-FWO), FWO Odysseus programme, Flanders Institute to encourage scientific and technological research in industry (IWT), Belgian Federal Science Policy Office (Belspo); University of Oxford, United Kingdom; Marsden Fund, New Zealand; Australian Research Council; Japan Society for Promotion of Science (JSPS); the Swiss National Science Foundation (SNSF), Switzerland; National Research Foundation of Korea (NRF); Villum Fonden, Danish National Research Foundation (DNRF), Denmark.

The VERITAS Collaboration acknowledges the support of the
U.S. Department of Energy Office of Science, the U.S. National Science
Foundation and the Smithsonian Institution, and NSERC in Canada. We
also acknowledge the excellent work of the technical support staff at
the Fred Lawrence Whipple Observatory and at the collaborating
institutions in the construction and operation of the instrument.

The MAGIC collaboration would like to thank
the Instituto de Astrof\'{\i}sica de Canarias
for the excellent working conditions
at the Observatorio del Roque de los Muchachos in La Palma.
The financial support of the German BMBF and MPG,
the Italian INFN and INAF,
the Swiss National Fund SNF,
the he ERDF under the Spanish MINECO
(FPA2015-69818-P, FPA2012-36668, FPA2015-68278-P,
FPA2015-69210-C6-2-R, FPA2015-69210-C6-4-R,
FPA2015-69210-C6-6-R, AYA2013-47447-C3-1-P,
AYA2015-71042-P, ESP2015-71662-C2-2-P, CSD2009-00064),
and the Japanese JSPS and MEXT
is gratefully acknowledged.
This work was also supported
by the Spanish Centro de Excelencia ``Severo Ochoa''
SEV-2012-0234 and SEV-2015-0548,
and Unidad de Excelencia ``Mar\'{\i}a de Maeztu'' MDM-2014-0369,
by grant 268740 of the Academy of Finland,
by the Croatian Science Foundation (HrZZ) Project 09/176
and the University of Rijeka Project 13.12.1.3.02,
by the DFG Collaborative Research Centers SFB823/C4 and SFB876/C3,
and by the Polish MNiSzW grant 745/N-HESS-MAGIC/2010/0.

\section{Appendix}
\label{appendixI}

List of sources used by NToO for IC-2012  season (Table~\ref{tab:sources2012-2013}) and  for IC-2013 and IC-2014  season (Table~\ref{tab:sources2014-2015}). In the  table the source name, the declination (DEC), the right ascension (RA), search bin radius, and threshold for sending alerts is listed.  The last column indicates
if the source belongs only to the MAGIC list or VERITAS list
or if the source is present  in the list for both experiments (BOTH).

\begin{longtable}{|c|c|c|c|c|c|c |}
\hline
L.P. & Source &   DEC & RA   &  Search radius & Threshold &  Exper.\\
& & (deg)  & (deg) & (deg) & ($\sigma$) & \\
\hline

1 & PMN J0948$+$ 0022&  0.3740&  147.2390&  1.21&  3.63&  VERITAS \\
2 &  BL 0414$+$ 009&  1.0900&  64.2187&  1.23&  3.16&  BOTH\\
3 & PKS B0906$+$ 015&  1.3600&  137.2920&  1.23&  3.63&  VERITAS\\
4 & RGB J0152$+$ 017&  1.7779&  28.1396&  1.24&  3.63&  VERITAS\\
5 & 3C~273&  2.0525&  187.2779&  1.25&  3.16&  BOTH\\
6 & BL 0323$+$ 022&  2.4208&  51.5583&  1.26&  3.16&  MAGIC\\
7 & MG1 J050533$+$ 0415&  4.2650&  76.3950&  1.30&  3.63&  VERITAS\\
8 & J123939$+$ 044409&  4.7000&  189.9000&  1.31&  3.63&  VERITAS\\
9 & HESS J0632$+$ 057&  5.8056&  98.2429&  1.34&  3.63&  VERITAS\\
10 & 1 ES 1212$+$ 078&  7.5347&  183.7958&  1.38&  3.16&  MAGIC\\
11 & 4C $+$ 09.57&  9.6300&  267.8900&  1.43&  3.63&  VERITAS\\
12 & PKS 0754$+$ 100&  9.9400&  119.3100&  1.44&  3.63&  VERITAS\\
13 & PKS 1502$+$ 106&  10.4940&  226.1040&  1.46&  3.63&  VERITAS\\
14 & CGRaBS J0211$+$ 1051&  10.8600&  32.8050&  1.46&  3.63&  VERITAS\\
15 & PKS 2032$+$ 107&  11.0000&  308.8600&  1.47&  3.63&  VERITAS\\
16 & PG 1553$+$ 113&  11.1900&  238.9292&  1.47&  3.16&  BOTH\\
17 & RGB 0847$+$ 115&  11.56389&  131.8038&  1.48&  3.16&  MAGIC\\
18 & CTA 102&  11.7310&  338.1520&  1.48&  3.63&  VERITAS\\
19 & BL 1722$+$ 119&  11.8708&  261.2679&  1.49&  3.16&  MAGIC\\
20 & 1ES 1440$+$ 122&  12.0111&  220.7010&  1.49&  3.63&  VERITAS\\
21 & M87&  12.3975&  187.6970&  1.50&  3.63&  VERITAS\\
22 & PKS 0528$+$ 134&  13.5320&  82.7350&  1.53&  3.63&  VERITAS\\
23 & 4C 14.23&  14.4200&  111.3200&  1.55&  3.63&  VERITAS\\
24 & RGB 0648$+$ 151&  15.2736&  102.1983&  1.57&  3.16&  BOTH\\
25 & 4c15.54&  15.8594&  241.7775&  1.58&  3.16&  MAGIC\\
26 & 3C 454.3&  16.1480&  343.4910&  1.59&  3.63&  VERITAS\\
27 & AO 0235$+$ 164&  16.6164&  39.6621&  1.60&  3.16&  BOTH\\
28 & RGB 0250$+$ 172&  17.2025&  42.6579&  1.61&  3.16&  MAGIC\\
29 & PKS 0735$+$ 178&  17.7053&  114.5308&  1.62&  3.16&  MAGIC\\
30 & OX 169&  17.7300&  325.8980&  1.63&  3.63&  VERITAS\\
31 & PKS 1717$+$ 177&  17.7517&  259.8042&  1.63&  3.16&  BOTH\\
32 & HB89 0317$+$ 185&  18.7594&  49.9658&  1.65&  3.16&  BOTH\\
33 & MG2 J071354$+$ 1934&  19.5830&  108.4820&  1.67&  3.63&  VERITAS\\
34 & 1ES 1741$+$ 196&  19.5858&  265.9908&  1.67&  3.16&  BOTH\\
35 & OJ 287&  20.1108&  133.7033&  1.68&  3.16&  BOTH\\
36 & RGB 1117$+$ 202&  20.2356&  169.2758&  1.68&  3.16&  MAGIC\\
37 & 1ES 0229$+$ 200&  20.2881&  38.2025&  1.68&  3.16&  BOTH\\
38 & RGB 0521$+$ 211&  21.2142&  80.4412&  1.71&  3.16&  BOTH\\
39 & PKS1222$+$ 21&  21.3794&  186.2270&  1.71&  3.16&  BOTH\\
40 & Crab Pulsar&  22.0140&  83.6330&  1.73&  3.63&  VERITAS\\
41 & RGB 0909$+$ 231&  23.1867&  137.2529&  1.75&  3.16&  MAGIC\\
42 & RGB 0321$+$ 236&  23.6031&  50.5000&  1.76&  3.16&  MAGIC\\
43 & PG 1424$+$ 240&  23.8000&  216.7517&  1.77&  3.16&  BOTH\\
44 & 1ES 1255$+$ 244&  24.2111&  194.3829&  1.77&  3.16&  MAGIC\\
45 & 0827$+$ 243&  24.2200&  127.4900&  1.77&  3.63&  VERITAS\\
46 &  1ES 0647$+$ 250&  25.0500&  102.6938&  1.79&  3.16&  MAGIC\\
47 & RGB 1417$+$ 257&  25.7236&  214.4858&  1.80&  3.16&  MAGIC\\
48 & W Comae&  28.2331&  185.3821&  1.86&  3.16&  BOTH\\
49 & Ton 599&  29.2460&  179.8830&  1.88&  3.63&  VERITAS\\
50 & HB89 0912$+$ 293&  29.5567&  138.9683&  1.89&  3.16&  MAGIC\\
51 & ON 325&  30.1169&  184.4671&  1.90&  3.16&  BOTH\\
52 & 1ES 1218$+$ 304&  30.1769&  185.3413&  1.90&  3.16&  BOTH\\
53& B2 1520$+$ 31&  31.7370&  230.5420&  1.94&  3.63&  VERITAS\\
54 & 4C 31.03&  32.1380&  18.2100&  1.95&  3.63&  VERITAS\\
55 & CGRaBS J1848$+$ 3219&  32.3170&  282.0920&  1.95&  3.63&  VERITAS\\
56 & B2 0619$+$ 33&  33.4360&  95.7180&  1.97&  3.63&  VERITAS\\
57 & HB89 1721$+$ 343&  34.2994&  260.8367&  1.99&  3.16&  MAGIC\\
58 & 1ES 0120$+$ 340&  34.3472&  20.7867&  1.99&  3.16&  MAGIC\\
59 & B2 2308$+$ 34&  34.4200&  347.7720&  1.99&  3.63&  VERITAS\\
60 & RGB 0706$+$ 377&  37.7433&  106.6321&  2.06&  3.16&  MAGIC\\
61 & NVSS 232914$+$ 3754&  37.9042&  352.309167&  2.06&  3.16&  MAGIC\\
62 & 1633$+$ 382&  38.1350&  248.8150&  2.06&  3.63&  VERITAS\\
63 & Mkn 421&  38.2089&  166.1138&  2.07&  3.16&  BOTH\\
64 & B3 2247$+$ 381&  38.4103&  342.5238&  2.07&  3.16&  BOTH\\
65 & RGB 0136$+$ 391&  39.1000&  24.1363&  2.08&  3.16&  MAGIC\\
66 & 0FGL J1641.4$+$ 3939&  39.6660&  250.3550&  2.09&  3.63&  VERITAS\\
67 & Mkn 501&  39.7603&  253.4675&  2.10&  3.16&  BOTH\\
68 & IC 310&  41.3247&  49.1792&  2.12&  3.63&  VERITAS\\
69 & TeV J2032$+$ 4130&  41.5100&  308.0830&  2.13&  3.63&  VERITAS\\
70 & NGC1275&  41.5117&  49.9504&  2.13&  3.16&  BOTH\\
71 & 1ES 2321$+$ 419&  42.1831&  350.9671&  2.14&  3.16&  BOTH\\
72 & BL Lac&  42.2778&  330.6804&  2.14&  3.16&  BOTH\\
73 & B3 0814$+$ 425&  42.3800&  124.5500&  2.14&  3.63&  VERITAS\\
74 & 1ES 1426$+$ 428&  42.6725&  217.1358&  2.15&  3.16&  BOTH\\
75 & 3C66A&  43.0356&  35.6650&  2.16&  3.16&  BOTH\\
76 & B3 1307$+$ 433&  43.0847&  197.3563&  2.16&  3.16&  MAGIC\\
77 & B3 1708$+$ 433&  43.3120&  257.4210&  2.16&  3.63&  VERITAS\\
78 & MG4J200112$+$ 4352&  43.8814&  300.3038&  2.17&  3.16&  BOTH\\
79 & B3 1343$+$ 451&  44.8830&  206.3880&  2.19&  3.63&  VERITAS\\
80 & GB6 B1310$+$ 4844&  48.4750&  198.1810&  2.25&  3.63&  VERITAS\\
81 & 1ES 1011$+$ 496&  49.4336&  153.7675&  2.26&  3.16&  BOTH\\
82 & 1150$+$ 497&  49.5190&  178.3520&  2.27&  3.63&  VERITAS\\
83 & 1ES 0927$+$ 500&  49.8406&  142.6567&  2.27&  3.16&  MAGIC\\
84 & BL 1ZW187&  50.2194&  262.0775&  2.28&  3.16&  MAGIC\\
85 & 1ES 1028$+$ 511&  50.8933&  157.8271&  2.28&  3.16&  MAGIC\\
86 & 1ES 2344$+$ 514&  51.7050&  356.7700&  2.30&  3.16&  BOTH\\
87 & 1ES 0806$+$ 524&  52.3000&  122.4542&  2.31&  3.16&  BOTH\\
88 & BZU J0742$+$ 5444&  54.7400&  115.6660&  2.34&  3.63&  VERITAS\\
89 & 4C55.17&  55.3828&  149.4092&  2.35&  3.16&  MAGIC\\
90 & RGB 1903$+$ 556&  55.6772&  285.7983&  2.36&  3.16&  MAGIC\\
91 & RGB 1058$+$ 564&  56.4697&  164.6570&  2.37&  3.16&  BOTH\\
92 & RBS 1409&  56.6569&  219.2404&  2.37&  3.16&  MAGIC\\
93 & PG 1246$+$ 586&  58.3414&  192.0783&  2.39&  3.16&  MAGIC\\
94 & exo 0706$+$ 5913&  59.1389&  107.6250&  2.40&  3.16&  BOTH\\
95 & 1ES 0033$+$ 595&  59.8347&  8.9692&  2.41&  3.16&  MAGIC\\
96 & S4 1030$+$ 61&  60.8520&  158.4640&  2.42&  3.63&  VERITAS\\
97 & RGB 0505$+$ 612&  61.2267&  76.4950&  2.43&  3.16&  MAGIC\\
98 & LSI $+$ 61 303&  61.2290&  40.1310&  2.43&  3.63&  VERITAS\\
99 & 1ES 1959$+$ 650&  65.1486&  299.9992&  2.47&  3.16&  BOTH\\
100 & S4  0954$+$ 658&  65.5653&  149.6967&  2.48&  3.16&  MAGIC\\
101 & CGRaBS J1849$+$ 6705&  67.0950&  282.3170&  2.49&  3.63& VERITAS\\
102 & RGB 1136$+$ 676&  67.6178&  174.1254&  2.49&  3.16&  MAGIC\\
103 & 1ES 0502$+$ 675&  67.6233&  76.9842&  2.49&  3.16&  BOTH\\
104 & GB6 J1700$+$ 6830&  68.5020&  255.0390&  2.50&  3.63&  VERITAS\\
105 & HB89 1749$+$ 701&  70.09750&  267.1367&  2.52&  3.16&  BOTH\\
106 & Mkn 180&  70.1575&  174.1100&  2.52&  3.16&  BOTH\\
107 & S5 0836$+$ 71&  70.8950&  130.3520&  2.52&  3.63&  VERITAS\\
108 & S5 0716$+$ 714&  71.3433&  110.4725&  2.53&  3.16&  BOTH\\
109 & S5 1803$+$ 78&  78.4680&  270.1900&  2.57&  3.63&  VERITAS\\
\hline

%\end{longtable}
%\vspace{0.5 cm}
\caption{List of sources used by NToO from November 2013 to December 2015. In the  table the source name, the declination (DEC), the right ascension (RA), search bin radius, and threshold for sending alerts is listed.  The last column indicates
if the source belongs only to the MAGIC list or VERITAS list
or if the source is present  in the list for both experiments (BOTH). }
\label{tab:sources2012-2013}	
\end{longtable} 

\begin{longtable}{|c|c|c|c|c|c|c |}
\hline
L.P. & Source &   DEC & RA   &  Search radius & Threshold &  Exper.\\
& & (deg)  & (deg) & (deg) & ($\sigma$) & \\
\hline
%1&  PG 1553+113 &  11.1902 &  238.9418 &  1.47 & 3.16  &  BOTH  \\
%2&  PKS 1717+177 & 17.7425 & 259.8300 & 1.63 & 3.16  & BOTH \\\
%3&  RBS 0413 & 18.8266 & 49.9094 & 1.65& 3.16  & BOTH  \\
%4&  RBS 0958 & 20.2269 & 169.3050 & 1.68 & 3.16  & MAGIC \\
%5&  MS 1458.8+2249 & 22.6388 & 225.2749 & 1.74 & 3.16  & MAGIC \\
%6&  PKS 1424+240 & 23.9750 & 216.7597 & 1.77 & 3.16  & BOTH   \\
%7&  W Comae &  28.2391 & 185.3740 & 1.86 & 3.16  & BOTH \\
%8&  1ES 1215+303 & 30.1093 & 184.4672 & 1.90 & 3.16  & BOTH \\
%9&  Mkn 421 & 38.2134 & 166.1199 & 2.07 & 3.16  & BOTH \\
%10&  Mkn 501 & 39.7631 & 253.4814 & 2.10 & 3.16  & BOTH  \\
%11&  1ES 2321+419 & 42.2001 & 350.9539 & 2.14 & 3.16  & BOTH \\
%12&  3C 66A & 43.0358 & 35.6617 & 2.16 & 3.16  & BOTH \\
%13&  GB6 J1838+4802 & 47.9939 & 279.6958 & 2.24 & 3.16  & MAGIC \\
%14&  TXS 1055+567 & 56.4801 & 164.6656 & 2.37 & 3.16  & BOTH \\
%15&  GB6 J1542+6129 & 61.4887 & 235.7294 & 2.43 & 3.16  & MAGIC \\
%16&  1ES 1959+650 & 65.1572 & 300.0204 & 2.47 & 3.16  & BOTH \\
%17&  S5 0716+71 & 71.3496 & 110.4757 & 2.53 & 3.16  & BOTH \\
%18&  RX J0805.4+7534 &  75.5878 & 121.3421 & 2.56 & 3.16  & MAGIC \\
1 & PG 1553+113& 11.1902& 238.9418& 1.47& 3.16 & BOTH \\
2 & PKS 1424+240&23.9750&216.7597&1.77& 3.16 &BOTH \\
3 &PKS 1717+177&17.7425&259.8300&1.63& 3.16 &BOTH   \\
4 &RBS 0413&18.8266&49.9094&1.65& 3.16 &BOTH      \\
5 &RBS 0958&20.2269&169.3050&1.68& 3.16 &MAGIC     \\
6 &RX J0805.4+7534& 75.5878&121.3421&2.56& 3.16 &MAGIC\\
7 &S5 0716+71&71.3496&110.4757&2.53& 3.16 & BOTH       \\
8 &TXS 1055+567&56.48010&164.6656&2.37& 3.16 &BOTH     \\
9 &W Comae& 28.2391&185.3740&1.86& 3.16 &BOTH          \\
10 &1ES 1215+303&30.1093&184.4672&1.90& 3.16 &BOTH   \\
11 &1ES 1959+650&65.1572&300.0204&2.47& 3.16 &BOTH         \\
12 &1ES 2321+419&42.2001&350.9539&2.14& 3.16 &BOTH    \\
13 &3C 66A&43.0358&35.6617&2.16& 3.16 &BOTH            \\
14 &GB6 J1542+6129&61.4887&235.7294&2.43& 3.16 &MAGIC   \\
15 &GB6 J1838+4802&47.9939&279.6958&2.24& 3.16 &MAGIC   \\
16 &MS 1458.8+2249&22.6388&225.2749&1.74& 3.16 &MAGIC   \\
17 &Mkn 421&38.2134&166.1199&2.07& 3.16 &BOTH            \\
18 &Mkn 501&39.7631&253.4814&2.10& 3.16 &BOTH          \\
19 &PMN J0948+0022&0.3740&147.2390&1.21& 3.63 &VERITAS   \\
20 &BL 0414+009&1.0900&64.2188&1.23& 3.63 &VERITAS       \\
21 &PKS B0906+015&1.3600&137.2920&1.23& 3.63 &VERITAS     \\
22 &RGB J0152+017&1.7779&28.1396&1.24& 3.63 &VERITAS      \\
23 &3C 273&2.0525&187.2779&1.25& 3.63 &VERITAS              \\
24 &MG1 J050533+0415&4.2650&76.3950&1.30& 3.63 &VERITAS  \\
25 &J123939+044409&4.7000&189.9000&1.31& 3.63 &VERITAS     \\
26 &HESS J0632+057&5.8056&98.2429&1.34& 3.63 &VERITAS    \\
27 &4C +09.57&9.6300&267.8900&1.43& 3.63 &VERITAS          \\
28 &PKS 0754+100&9.9400&119.3100&1.44& 3.63 &VERITAS      \\
29 &PKS 1502+106&10.4940&226.1040&1.45& 3.63 &VERITAS     \\
30 &CGRaBS J0211+1051&10.8600&32.8050&1.47& 3.63 &VERITAS \\
31 &PKS 2032+107&11.0000&308.8600&1.47& 3.63 &VERITAS     \\
32 &CTA 102&11.7310&338.1520&1.48& 3.63 &VERITAS          \\
33 &1ES 1440+122&12.0111&220.7010&1.49& 3.63 &VERITAS     \\
34 &M 87&12.3975&187.6970&1.50& 3.63 &VERITAS              \\
35 &PKS 0528+134&13.5320&82.7350&1.53& 3.63 &VERITAS       \\
36 &4C 14.23&14.4200&111.3200&1.55& 3.63 &VERITAS         \\
37 &RGB 0648+151&15.2736&102.1983&1.57& 3.63 &VERITAS    \\
38 &3C 454.3&16.1480&343.4910&1.59& 3.63 &VERITAS         \\
39  &AO 0235+164&16.616&39.6621&1.60& 3.63 &VERITAS       \\
40 &OX 169&17.7300&325.8980&1.63& 3.63 &VERITAS           \\
41 &MG2 J071354+1934&19.5830&108.4820&1.67& 3.63 &VERITAS \\
42 &1ES 1741+196&19.5858&265.9908&1.67& 3.63 &VERITAS     \\
43 &OJ 287&20.1108&133.7033&1.68& 3.63 &VERITAS         \\
44 &1ES 0229+200&20.2881&38.2025&1.69& 3.63 &VERITAS  \\
45 &RGB 0521+211&21.2142&80.4413&1.71& 3.63 &VERITAS\\
46 &PKS1222+21&21.3794&186.2271&1.71& 3.63 &VERITAS\\
47 &Crab Pulsar&22.0140&83.6330&1.72& 3.63 &VERITAS\\
48 &0827+243&24.2200&127.4900&1.77& 3.63 &VERITAS\\
49 &Ton 599&29.2460&179.8830&1.88& 3.63 &VERITAS\\
50 &1ES 1218+304&30.1769&185.3413&1.90& 3.63 &VERITAS\\
51 &B2 1520+31&31.7370&230.5420&1.94& 3.63 &VERITAS\\
52 &4C 31.03&32.1380&18.2100&1.94& 3.63 &VERITAS   \\
53 &CGRaBS J1848+3219&32.3170&282.0920&1.95& 3.63 &VERITAS\\
54 &B2 0619+33&33.4360&95.7180&1.97& 3.63 &VERITAS    \\
55 &B2 2308+34&34.4200&347.7720&1.99& 3.63 &VERITAS     \\
56 &1633+382&38.1350&248.8150&2.06& 3.63 &VERITAS         \\
57 &B3 2247+381&38.4103&342.5238&2.07& 3.63 &VERITAS     \\
58 &0FGL J1641.4+3939&39.6660&250.3550&2.09& 3.63 &VERITAS\\
59 &IC 310&41.3247&49.1792&2.12& 3.63 &VERITAS    \\
60 &TeV J2032+4130&41.5100&308.0830&2.13& 3.63 &VERITAS\\
61 &NGC1275&41.5117&49.9504&2.13& 3.63 &VERITAS  \\
62 &BLLac&42.2778&330.6804&2.14& 3.63 &VERITAS\\
63 &B3 0814+425&42.3800&124.5500&2.14& 3.63 &VERITAS\\
64 &1ES 1426+428&42.6725&217.1358&2.15& 3.63 &VERITAS\\
65 &B3 1708+433&43.3120&257.4210&2.16& 3.63 &VERITAS\\
66 &MG4J200112+4352&43.8814&300.3038&2.17& 3.63 &VERITAS\\
67 &B3 1343+451&44.8830&206.3880&2.19& 3.63 &VERITAS\\
68 &GB6 B1310+4844&48.4750&198.1810&2.25& 3.63 &VERITAS\\
69 &1ES 1011+496&49.4336&153.7675&2.26& 3.63 &VERITAS\\
70 &1150+497&49.5190&178.3520&2.26& 3.63 &VERITAS\\
71 &1ES 2344+514&51.7050&356.7700&2.30& 3.63 &VERITAS\\
72 &1ES 0806+524&52.3000&122.4542&2.31& 3.63 &VERITAS\\
73 &BZU J0742+5444&54.7400&115.6660&2.34& 3.63 &VERITAS\\
74 &exo 0706+5913&59.1389&107.6250&2.40& 3.63 &VERITAS\\
75 &S4 1030+61&60.8520&158.4640&2.42& 3.63 &VERITAS\\
76 &LSI +61 303&61.2290&40.1310&2.43& 3.63 &VERITAS\\
77 &CGRaBS J1849+6705&67.0950&282.3170&2.49& 3.63 &VERITAS\\
78 &1ES 0502+675&67.6233&76.9842&2.49& 3.63 &VERITAS\\
79 &GB6 J1700+6830&68.5020&255.0390&2.50& 3.63 &VERITAS\\
80 &HB89 1749+701&70.0975&267.1367&2.52& 3.63 &VERITAS \\
81&Mkn 180&70.1575&174.1100&2.52& 3.63 &VERITAS  \\
82&S5 0836+71&70.8950&130.3520&2.52& 3.63 &VERITAS\\
83&S5 1803+78&78.4680&270.1900&2.57& 3.63 &VERITAS \\

\hline
\caption{List of sources used by NToO for IC-2013 and IC-2014 season. }
\label{tab:sources2014-2015}	
\end{longtable} 

\end{document}